\documentclass[a4paper,fleqn,usenatbib,usedcolumn]{mnras}
\pdfoutput=1 % for arxiv
\usepackage{mathptmx}
\usepackage[T1]{fontenc}
\usepackage{ae,aecompl}

\usepackage{graphicx}	% Including figure files
\usepackage{amsmath}	% Advanced maths commands
\usepackage{amssymb}	% Extra maths symbols

\usepackage{color}
\usepackage{xfrac}
\usepackage{algorithm2e}
\usepackage{multicol}
\usepackage{multirow}

\pdfpageattr{/Group <</S /Transparency /I true /CS /DeviceRGB>>}

\definecolor{Gray}{RGB}{148,150,152}
\definecolor{Plum}{RGB}{146,38,143}
\definecolor{Bittersweet}{RGB}{192,79,23}

\newcommand{\forbidden}[3]{\mbox{[#1\,\textsc{#2}]{\smaller#3}}}
\newcommand{\Halpha}{\mbox{H\ensuremath{\alpha}}}
\newcommand{\Hbeta}{\mbox{H\ensuremath{\beta}}}
\newcommand{\Hgamma}{\mbox{H\ensuremath{\gamma}}}
\newcommand{\angstrom}{\textup{\AA}}

\def\aj{AJ}% Astronomical Journal
\def\araa{ARA\&A}% Annual Review of Astron and Astrophys
\def\apj{ApJ}% Astrophysical Journal
\def\apjl{ApJ}% Astrophysical Journal, Letters
\def\apjs{ApJS}% Astrophysical Journal, Supplement
\def\apss{Ap\&SS}% Astrophysics and Space Science
\def\aap{A\&A}% Astronomy and Astrophysics
\def\mnras{MNRAS}% Monthly Notices of the RAS
\def\nat{Nature}% Nature
\title[Forward modelling of metallicity gradients]{Inferring gas-phase metallicity gradients of galaxies at the seeing limit: A forward modelling approach}

\author[David Carton et al.]{
David Carton$^{1,2}$\thanks{E-mail:  \href{mailto:david.carton@univ-lyon1.fr}{david.carton@univ-lyon1.fr}},
Jarle Brinchmann$^{1,3}$,
Maryam Shirazi$^{1,4}$,
Thierry Contini$^{5,6}$,\newauthor{}
Beno\^{i}t Epinat$^{5,6,7}$,
Santiago Erroz-Ferrer$^{4}$,
Raffaella A. Marino$^{4}$,\newauthor{}
Thomas P. K. Martinsson$^{1,8,9}$,
Johan Richard$^{2}$,
Vera Patr\'{i}cio$^{2}$,
\\
$^{1}$Leiden Observatory, Leiden University, PO Box 9513, 2300 RA, Leiden, The Netherlands\\
$^{2}$Univ Lyon, Univ Lyon1, Ens de Lyon, CNRS, Centre de Recherche Astrophysique de Lyon UMR5574, 69230, Saint-Genis-Laval, France\\
$^{3}$Instituto de Astrof\'{i}sica e Ci\^{e}ncias do Espa\c{c}o, Universidade do Porto, CAUP, Rua das Estrelas, 4150-762, Porto, Portugal\\
$^{4}$Institute for Astronomy, ETH Z\"{u}rich, Wolfgang-Pauli-Str. 27, 8093, Z\"{u}rich, Switzerland\\
$^{5}$IRAP, Institut de Recherche en Astrophysique et Plan\'{e}tologie, CNRS, 14, avenue Edouard Belin, 31400, Toulouse, France\\
$^{6}$Universit\'{e} de Toulouse, UPS-OMP, Toulouse, France\\
$^{7}$Aix Marseille Univ, CNRS, LAM, Laboratoire d'Astrophysique de Marseille, Marseille, France\\
$^{8}$Instituto de Astrof\'{i}sica de Canarias (IAC), 38205, La Laguna, Tenerife, Spain\\
$^{9}$Departamento de Astrof\'{i}sica, Universidad de La Laguna, 38206, La Laguna, Tenerife, Spain\\
}

\date{Accepted 2017 February 28. Received 2017 February 28; in original form 2016 June 2}

\pubyear{2017}

\begin{document}
\label{firstpage}
\pagerange{\pageref{firstpage}--\pageref{lastpage}}
\maketitle

%\begin{abstract}
\begin{abstract}
We present a method to recover the gas-phase metallicity gradients from integral field spectroscopic (IFS) observations of barely resolved galaxies.
We take a forward modelling approach and compare our models to the observed spatial distribution of emission line fluxes, accounting for the degrading effects of seeing and spatial binning.
The method is flexible and is not limited to particular emission lines or instruments.
We test the model through comparison to synthetic observations and use downgraded observations of nearby galaxies to validate this work.
As a proof of concept we also apply the model to real IFS observations of high-redshift galaxies.
From our testing we show that the inferred metallicity gradients and central metallicities are fairly insensitive to the assumptions made in the model and that they are reliably recovered for galaxies with sizes approximately equal to the half width at half maximum of the point-spread function.
However, we also find that the presence of star forming clumps can significantly complicate the interpretation of metallicity gradients in moderately resolved high-redshift galaxies.
Therefore we emphasize that care should be taken when comparing nearby well-resolved observations to high-redshift observations of partially resolved galaxies.
\end{abstract}

%\end{abstract}


%%%%%%%%%%%%%%%%%%%%%%%%%%%%%%%%%%%%%%%%%%%%%%%%%%

%%%%%%%%%%%%%%%%% BODY OF PAPER %%%%%%%%%%%%%%%%%%

\section{Introduction}

It is well known that star forming galaxies present a moderately tight relation between their stellar masses and their star formation rates \citep[e.g.][]{2004MNRAS.351.1151B,2007ApJ...660L..43N,2014ApJ...795..104W}.
Further it has been well established that the star formation rates of these galaxies is correlated with their gas content \citep[e.g.][]{1998ApJ...498..541K,2008AJ....136.2846B,2010MNRAS.407.2091G}, but that these gas reservoirs are insufficient to sustain star formation periods $> 0.7\,\textrm{Gyr}$ \citep{2013ApJ...768...74T}.
It has been suggested that galaxies grow in a regulated fashion which maintains an equilibrium between these quantities, where the star formation rate is limited by the supply and removal of gas (inflows/outflows) \citep{2010ApJ...718.1001B,2012MNRAS.421...98D,2013ApJ...772..119L}.
Therefore to understand how galaxies form and evolve we should study gas flowing into and out from galaxies.

Gas-phase metallicity\footnote{Throughout this work we use metallicity, gas-phase metallicity and oxygen abundance, $12 + \log_{10}\left(\textrm{O}/\textrm{H}\right)$, interchangeably.} provides an indirect tracer of gas flows in galaxies.
While gas-phase metallicity does not directly track the volume of gas in a galaxy, it does, however, indicate the origin of the gas.
To understand this it is often helpful to consider metallicity in the context of two other fundamental observables: the star-formation rate, and the stellar mass.
Both gas-phase metallicity and stellar mass track a similar quantity, the time-integrated star-formation history.
However, the presence of gas flows will cause the metallicity and stellar mass to diverge from a simple one-to-one relation.

Inflows and outflows can both have similar effects, both lowering the observed metallicity, one introduces metal-poor gas into the system, whilst the other preferentially expels metals entrained in winds \citep[see][]{2005ARA&A..43..769V}.
Studying the interplay of the star formation rate, stellar mass, and gas-phase metallicity is imperative to understanding the relation to the regulated growth of galaxies \citep[e.g.][]{2013ApJ...772..119L, 2016MNRAS.456.2140M}.

By examining the metallicity gradients of massive ($\ga10^8\,\textrm{M}_{\sun{}}$) low-redshift galaxies it has been found that the centres of galaxies are more typically metal-rich than their outskirts \citep{1992MNRAS.259..121V,1994ApJ...420...87Z}.
Furthermore it is often claimed that when normalized for disc scale-length, the same (common) metallicity gradient is found in all isolated galaxies \citep{2014A&A...563A..49S,2015MNRAS.448.2030H}.
This is not, however, the case for interacting or non-isolated galaxies, for which the metallicity profiles are typically shallower \citep{2012ApJ...753....5R}.
In these cases \citet{2010ApJ...710L.156R} have suggested that galaxy-galaxy interactions have triggered strong radial flows of gas towards the galaxy centre which act to temporarily erase the common metallicity gradient.

There are numerous reports of high redshift ($z \ga 1$) galaxies having inverted (positive) metallicity gradients \citep[e.g.][]{2012A&A...539A..93Q,2013ApJ...765...48J,2016ApJ...820...84L}.
However, this phenomenon for galaxies to have central regions more metal poor than their outskirts is not normally observed in low redshift galaxies.
It has been suggested that anomalously metal-poor centres may be a result of low-metallicity gas being deposited in the inner regions of galaxies: either via cold flow accretion \citep[e.g.][]{2010Natur.467..811C, 2013MNRAS.435.2918M, 2014A&A...563A..58T} or the transport of gas from the outer disc \citep{2012A&A...539A..93Q}.
Support for these ideas comes with the indication that the metallicity gradient is correlated with the specific star-formation rate, with the trend for aggressively star-forming (starbursting) galaxies to possess flatter (less negative) or even positive metallicity gradients \citep{2014MNRAS.443.2695S}.
This could be consistent with low redshift results that interacting galaxies exhibit flatter metallicity gradients, since interacting galaxies often show elevated star formation activity.

Measuring the metallicity gradients of high-redshift galaxies is not straightforward as one has to contend with the effects of seeing \citep[e.g.][]{2014A&A...561A.129M}.
Observing strongly lensed galaxies has proven to be a successful approach for overcoming the loss of resolution \citep[e.g.][]{2013ApJ...767..106Y}.
However, with lensing alone it is hard to survey the larger galaxy population, and in particular assess environment effects. 
Therefore, as a complement, we should attempt to derive the metallicity gradients of barely resolved galaxies, correcting for the effects of seeing.
In recent surveys \citet{2014MNRAS.443.2695S} and \citet{2016ApJ...831..149W} use integral field spectroscopy (IFS) to provide metallicity gradients for a large sample of $0.6<z<2.6$ galaxies.
After measuring the seeing corrupted metallicity gradients they applied a correction factor to infer the true uncorrupted metallicity gradient.
Here we will present a similar, but inverse approach for deriving the true metallicity gradient in galaxies from IFS observations.
Instead of applying an a posteriori correction we propose a forward modelling approach in which we directly fit a model to the emission-line flux data.
From this model we can derive both the true metallicity gradient and its associated uncertainty.
Unlike previous methods, our approach is flexible and is not limited to a particular set of emission lines.
Our method can therefore be applied to galaxies observed over a variety of redshifts and/or with different instruments.

This paper is dedicated to outlining and testing a model which we shall apply in future work using the Multi Unit Spectroscopic Explorer (MUSE) \citep[][and in prep.]{2010SPIE.7735E..08B}.

We structure the paper as follows.
Section~\ref{sec:model_description} provides a detailed description of our method.
Afterwards we perform a comprehensive series of tests to analyse our model (Section~\ref{sec:model_testing}).
In Section~\ref{sec:application} we apply our method to real data and discuss some characteristics of the model.
Finally we summarize our results in Section~\ref{sec:conclusions}.
Throughout the paper we assume a \mbox{$\Lambda\textrm{CDM}$} cosmology with $H_0 = 70\ \textrm{km}\,\textrm{s}^{-1}\,\textrm{Mpc}^{-1}$, $\Omega_{\mathrm{m}} = 0.3$ and $\Omega_{\Lambda} = 0.7$.

\section{Model Description}\label{sec:model_description}

We are interested in measuring the metallicity gradients of distant galaxies.
However, our observations are often limited by the resolution of the telescope.
The point spread function (PSF) can have two effects on the metallicity gradient.
Firstly we expect that the larger the PSF, the flatter the observed metallicity gradient will be.
However, the PSF is also wavelength dependent and will alter the emission-line ratios and ultimately the derived metallicity in a complex manner.
Applying an a posteriori correction to infer the true metallicity gradient would be non-trivial. Here we present the opposite approach whereby we construct a model galaxy with a given metallicity profile and predict the 2D flux distribution.
We can fit the predicted fluxes to the observed fluxes and thereby find the best-fit metallicity gradient.
In this section we will describe this model and fitting procedure.

\subsection{Simulating Observations}\label{sec:model_outline}

We shall now outline the workflow that we use to simulate observations, i.e. how we project the model from the source plane to the observed flux.
At this point we will not concern ourselves with the physical properties (metallicity etc.) of the galaxy model itself.

To address the problem outlined above, our simulated observations must propagate the effects of seeing.
In addition, however, we must also mimic the aggregation (or ``binning'') of spaxels\footnote{spatial pixel}.
The binning of spaxels is often required to increase the signal-to-noise ratio (S/N) of the data, but at the cost of further spatial resolution loss.

\begin{figure}
\includegraphics[width=\linewidth]{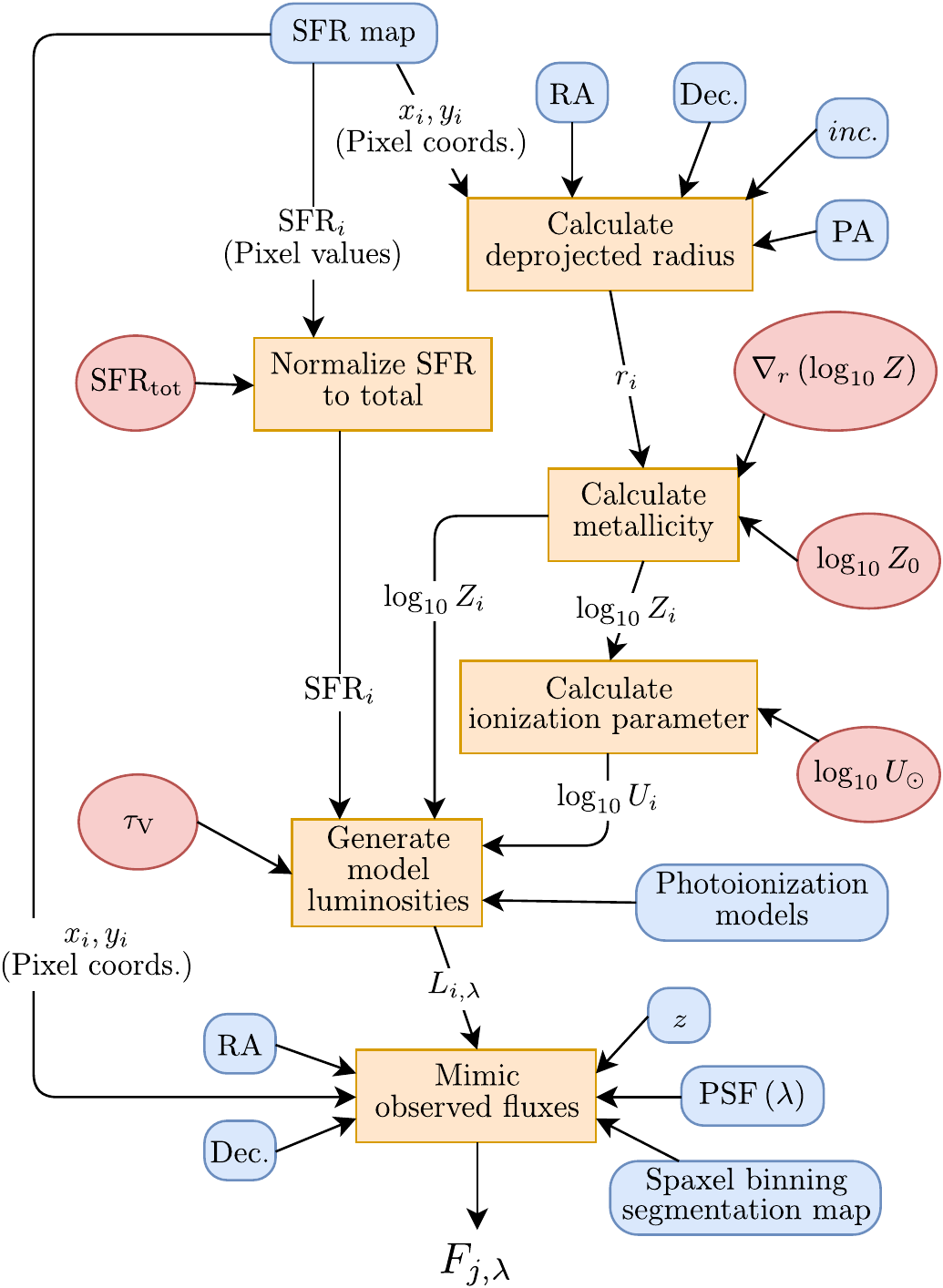}
\caption{Directed acyclic graph outlining the model workflow for generating model fluxes $F_{j,\lambda}$.
Fixed model inputs are represented as blue rectangles with rounded corners.
The five free parameters to the model are shown as red ellipses.
Computation steps within the model are drawn as orange rectangles.
$i$\textsuperscript{th} subscripts denote values assigned for each pixel in the input SFR map.
}
\label{fig:flowchart}
\end{figure}

We shall now describe our model.
To accompany this text we show a schematic outline of the model in Fig.~\ref{fig:flowchart}.
Our methodology is as follows:
\begin{enumerate}
\item \label{item:sfrmap}The galaxy is initialized from a star formation rate (SFR) map.
This map is a 2D Cartesian grid which lies in the plane of the sky.
For simplicity we treat each pixel to be represented by a point source situated at the centre of the pixel, and with a star formation rate (SFR) equal to that of the whole pixel.
In practice, to ensure the model is well-sampled, we will oversample our SFR maps by a factor two or three.

\item \label{item:galaxy_model} We use the galaxy model to associate a set of emission line luminosities to each point source.
We project each point source through the galaxy model (the galaxy lies in a plane inclined with respect to the observer).
Given the projected galaxy-plane coordinates and the SFR, the galaxy model generates a list of emission line fluxes as a function of position in the galaxy.
(The details of the galaxy model will be given in Section~\ref{sec:galaxy_model}).

\item \label{item:psf2pix} We now simulate image pixelization and PSF effects.
An output image pixel grid is constructed with same geometry as that of the observed image.
We calculate the distance from each point source to the centre of each pixel.
By evaluating the PSF at these distances we can approximate how much flux is diffused from each point source into each output pixel.

\item \label{item:pix2bin} To mimic the effects of aggregating spaxels together to increase the S/N, we also coadd the model pixels to match the exact binning that was applied to the data.
\end{enumerate}

In step~\ref{item:galaxy_model} we project source coordinates into the galaxy model plane.
This requires four morphological parameters: the Right Ascension (RA) and Declination (Dec.) of the galaxy centre, the inclination (inc.) of the galaxy, and the position angle (PA) of the major axis on the sky.
Partly for reasons of computational efficiency these morphological parameters are fixed a-priori.
The galaxy morphology can, for example, be determined from either high-resolution imaging or the kinematics of the ionized gas.
When fitting the model we will need to repeat steps~\ref{item:galaxy_model}--\ref{item:pix2bin} many times.
We can, however, vastly reduce the computation time if we cache the mapping operations (steps~\ref{item:psf2pix} \& \ref{item:pix2bin}) as a single sparse\footnote{The matrix is sparse as we only actually evaluate the PSF in step~\ref{item:psf2pix} for the closest pairs of point sources and  output pixels. The maximum evaluation distance is chosen to enclose 99.5\% of the PSF.} matrix.

So far we have only outlined how we simulate observations.
We have not yet touched upon how the emission-line luminosities are generated.
Our methodology divides this into two separate components: an SFR map and the galaxy model (i.e. steps~\ref{item:sfrmap} \& \ref{item:galaxy_model}, respectively).
Essentially, the former describes the 2D spatial emission-line intensity distribution, and the latter the 2D line-ratio distribution.
In the following sections we will describe both these components.

\subsection{Star Formation Rate (SFR) Maps}

Nebular emission lines are associated with the \ion{H}{ii} regions that surround young massive stars.
We therefore need to model the spatial SFR distribution.
The simplest approach would be to assume that the star formation rate density declines exponentially with radius, but while this might be an acceptable approximation, it is difficult for any parametric model to accurately describe the SFR distribution of a galaxy.
We shall later show that the clumpy nature of the SFR can have important consequences for the metallicity profile that we infer (see \S \ref{sec:real_tests}).
If a realistic (and reliable) empirical map of the SFR can be obtained then we should input this into the modelling.
In Appendix~\ref{sec:stochmod} we describe how these maps can be obtained in practice.
It is important to note that the map should have higher resolution than the data we are modelling.

The SFR map is not, however, entirely fixed a priori; to allow some flexibility in the model fit we shall allow one free parameter in the SFR.
We introduce a normalization constant, the total star formation rate ($\textrm{SFR}_\mathrm{tot}$) which is used to rescale the SFR map, and thereby it also rescales the emission-line luminosities without altering the line ratios in any way.

\subsection{The Galaxy Model}\label{sec:galaxy_model}

In our model we describe a galaxy as a series of \ion{H}{ii} regions, each with a SFR set by the input SFR map.
We assume the galaxy is infinitesimally thin, lying in an inclined plane.
Apart from the SFR distribution, the galaxy model is axisymmetric.
I.e. the emission line ratios only depend on one coordinate, $r$, the galactocentric radius.

There are three \ion{H}{ii} region properties in our model which set the observed line-ratios: metallicity, ionization parameter, and attenuation due to dust.
We shall now describe the radial parametrizations of these components.

\subsubsection{Metallicity and Ionization Parameter}

The physical properties of \ion{H}{ii} regions determine the observed emission-line intensities.
Varying elemental abundances alters the cooling rate of an \ion{H}{ii} region and thereby impacts upon the thermal balance of the \ion{H}{ii} region.
Temperature sensitive emission line ratios have long been used to infer the abundances of an \ion{H}{ii} region \citep{1959ApJ...130...45A}.
However, metallicity does not single-handedly control the emission-line intensities of \ion{H}{ii} regions.
Indeed the line-ratios will be affected by variations in the electron density and changes due to the ionizing continuum spectrum \citep{2013ApJ...774..100K}.
Theoretical photoionization models partly encapsulate these effects in the dimensionless ionization-parameter, $U$, which is in effect the ratio of the number density of ionizing photons to the number density of hydrogen atoms.
At fixed metallicity the largest variation in line ratios with physical properties is function of the ionization parameter \citep{2000ApJ...542..224D}.
So, similarly for our galaxy model we will assume that the emission line luminosities at each spatial position in the galaxy are prescribed by these two parameters: metallicity and ionization parameter.
We therefore need to parametrize both metallicity and ionization parameter spatially throughout the galaxy disc.

It has long been established that the metallicity in the inner disc of low redshift galaxies is well described by simple exponential function \citep[e.g.][]{2010ApJS..190..233M}.
With this precedent, and in accordance with others \citep[e.g.][]{2012A&A...539A..93Q}, we shall adopt the same functional form
\begin{equation}\label{eq:metallicity_profile}
\log_{10}Z(r) = \nabla_r\left(\log_{10}Z\right) r + \log_{10}Z_0,
\end{equation}
where $r$ is the radius, $\nabla_r\left(\log_{10}Z\right)$ is the metallicity gradient, and $\log_{10}Z_0$ is the metallicity at the galaxy centre.

\begin{figure}
\includegraphics[width=\linewidth]{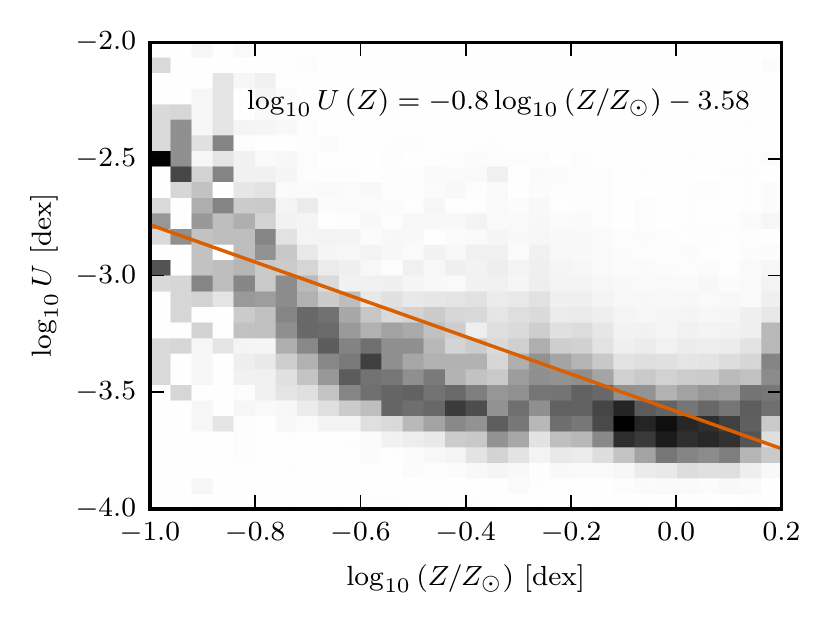}
\caption{Anti-correlation in the SDSS DR7 sample between ionization parameter, $\log_{10}U$, and central metallicity, $\log_{10}Z$.
SDSS galaxies show as a grey histogram.
The histogram is normalized per each metallicity bin (i.e. column).
The orange line indicates the best fit solution for the theoretical $U \propto Z^{-0.8}$ dependence.
To exclude active galactic nuclei (AGN) contamination we use the star-forming classification of \citet{2004MNRAS.351.1151B} (with a cut on emission-line $\textrm{S/N} > 10$).
To further exclude weak AGN we require that the stellar surface-mass density within the fibre is $< 10^{8.3}\,\textrm{M}_{\sun{}}/\textrm{kpc}^2$.
Note that because of the AGN removal our sample does not extend to very high metallicities.
}
\label{fig:SDSS_logZlogU}
\end{figure}

In contrast, the ionization parameter may depend on the local environmental conditions of the \ion{H}{ii} region, and therefore is not necessarily a simple function of galactocentric radius.
It would be very computationally challenging to non-parametrically incorporate the ionization parameter into the model.
We wish to have a simple one parameter description for the ionization parameter as a function of radius, but we do not wish to assume the ionization parameter to be constant throughout the galaxy.
Instead we exploit a natural anti-correlation between ionization parameter and metallicity \citep{1986ApJ...307..431D}.
The origin of this anti-correlation has been discussed fully in \citet{2006ApJ...647..244D}.
But to summarize, fewer ionizing photons escape from higher metallicity stars because at higher abundances stellar winds are more opaque and the photospheres scatter more photons.
These effects combined predict an anti-correlation between ionization parameter and metallicity with dependence $U \propto Z^{-0.8}$.
In Fig.~\ref{fig:SDSS_logZlogU} we show the dependence of ionization parameter on metallicity for the Sloan Digital Sky Survey (SDSS; \citet{2000AJ....120.1579Y}) Data Release 7 (DR7; \citet{2009ApJS..182..543A}).
It is clear that the SDSS sample broadly follows the  $U \propto Z^{-0.8}$, although at low metallicities ($\la -0.5\,\textrm{dex}$) the data implies a steeper dependence and is better described with a second-order polynomial.

In our galaxy model we shall couple the ionization parameter to the metallicity using
\begin{equation}\label{eq:ionization_parameter_coupling}
\log_{10}U\left(Z\right) = -0.8 \log_{10}\left(Z/Z_{\sun}\right) + \log_{10}U_{\sun},
\end{equation}
where $Z_{\sun}$ is solar abundance and $\log_{10}U_{\sun}$ is the ionization parameter at solar abundance.
We consider $\log_{10}U_{\sun}$ to be constant throughout the galaxy.
It has been suggested that higher redshift galaxies exhibit elevated ionization-parameters \citep{2014ApJ...787..120S,2015ApJ...812L..20K}, therefore we will allow the constant offset, $\log_{10}U_{\sun}$, to be a free parameter.

There is a second, but equally important reason for coupling the ionization-parameter to the metallicity.
In a typical use case of the model, we will have a galaxy with only a limited set of emission lines observed (e.g. \forbidden{O}{ii}{3727,3729}, \Hbeta{}, \forbidden{O}{iii}{5007}).
With these three emission lines the infamous $\textrm{R}_{23}$ degeneracy arises.
See for instance \citet{1991ApJ...380..140M} and \citet{2002ApJS..142...35K} who provide informative discussions of this degeneracy.
In this case, solving for metallicity produces two solutions, one low metallicity and the other high. 
Without additional information it is impossible to constrain which is the true solution.
However, consider the scenario in which we simultaneously measure a high $\textrm{O}_{32} = \left(\forbidden{O}{iii}{5007}/\forbidden{O}{ii}{3727,3729}\right)$ ratio, from this we would infer a high ionization-parameter.
By assuming metallicity and ionization-parameter are anti-correlated we could conclude the low-metallicity (high ionization-parameter) solution to be the correct one.
Our modelled galaxies therefore possess both metallicity and ionization parameter gradients, the slopes of which are anti-correlated with one another.

In this paper we adopt the photoionization models of \citet[][herein D13]{2013ApJS..208...10D}\defcitealias{2013ApJS..208...10D}{D13}.
In addition to metallicity and ionization parameter, these models introduce a third parameter, $\kappa$, that allows non-equilibrium electron energy distributions \citep{2012ApJ...752..148N}.
We will, however, limit ourselves to the traditional Maxwell-Boltzmann case \mbox{($\kappa=\infty$)}.
These photoionization models have been computed on a grid spanning \mbox{$0.05\,Z_{\sun} \le Z \le 5\,Z_{\sun}$}\footnote{The undepleted solar abundance of these photoionization models is $12 + \log_{10}\left(\textrm{O}/\textrm{H}\right) = 8.69$ \citep{2010Ap&SS.328..179G}.} and \mbox{$-3.98 \la \log_{10}U \la -1.98$}.
However, our above parametrization of $Z(r)$ and $\log_{10}U(r)$ is not explicitly bound to this region.
And since we do not wish to extrapolate the photoionization model grids, we ``clip'' $Z(r)$ and $\log_{10}U(r)$ so that they do not depart from the grid region.
I.e. where $Z(r) < 0.05\,Z_{\sun}$ we set $Z(r) = 0.05\,Z_{\sun}$ and likewise where $Z(r) > 5\,Z_{\sun}$ we set $Z(r) = 5\,Z_{\sun}$.
In Appendix~\ref{sec:model_line_ratios} we show the \citetalias{2013ApJS..208...10D} photoionization model grids for a few standard line-ratios.

The \citetalias{2013ApJS..208...10D} models adopt an electron density $n_\mathrm{e} \sim 10\,\textrm{cm}^{-3}$.
This is thought to be appropriate for low redshift galaxies, but this is not necessarily the case for high redshift ($z \ga 1$) galaxies \citep[e.g.][]{2014ApJ...787..120S, 2016ApJ...816...23S}.
We caution the reader that if our model is to be applied to high redshift galaxies, different photoionization models would likely be needed.
Indeed, the model could easily be extended to include the electron density of the galaxy as an additional free parameter.
However, since we will be applying this model to $z \la 1$ galaxies, we simply choose to fix the electron density at $n_\mathrm{e} \sim 10\,\textrm{cm}^{-3}$.

It is also worth noting that \citetalias{2013ApJS..208...10D} models assume that the underling stellar population has a continuous star formation history (as opposed to a instantaneous burst).
But, since we are applying our model to poorly resolved data, we are in effect averaging over many individual \ion{H}{ii} regions.
Therefore, while an instantaneous burst might be most appropriate for modelling individual \ion{H}{ii} regions, we consider the continuous star-formation assumption to be more valid for our purposes.

The line fluxes are scaled to luminosities based on the SFR map, with the following scaling relation between \Halpha{} luminosity and SFR as taken from \citet{1998ARA&A..36..189K}
\begin{equation}\label{eq:halpha_sfr_law}
\frac{L(\textrm{H}\alpha)}{\mathrm{erg}\,\mathrm{s}^{-1}} = \frac{1}{7.9 \times 10^{-42}} \frac{\textrm{SFR}}{\textrm{M}_{\sun}\,\mathrm{yr}^{-1}}.
\end{equation}
This assumes a \citet{1955ApJ...121..161S} initial mass function, consistent with the \citetalias{2013ApJS..208...10D} photoionization modelling.

The emission-line luminosities are computed as follows:
\begin{enumerate}
\item Evaluate the metallicity for each radial coordinate using equation~\ref{eq:metallicity_profile} (for given values of $\log_{10}Z_0$ and $\nabla_r\left(\log_{10}Z\right)$).

\item Clip $\log_{10}Z(r)$ to the metallicity range of the photoionization model grid.

\item Calculate the associated ionization parameter using equation~\ref{eq:ionization_parameter_coupling} (for a given value of $\log_{10}U_{\sun}$).

\item Clip $\log_{10}U(r)$ to the ionization parameter range of the photoionization models.

\item Infer the relative emission line luminosities by interpolating the photoionization grid at $\left(\log_{10}Z(r), \log_{10}U(r)\right)$.

\item Scale the emission-line luminosities appropriate for the $\textrm{SFR}$ using equation~\ref{eq:halpha_sfr_law}.
\end{enumerate}

\subsubsection{Dust attenuation}

There remains one hitherto undiscussed ingredient in the model, the attenuation due to dust.
Since dust attenuation is wavelength dependent it will alter the emission-line ratios.

We adopt the dust absorption curve appropriate for \ion{H}{ii} regions as proposed by \citet{2000ApJ...539..718C} 
\begin{equation}
L_\mathrm{ext}(\lambda) = L(\lambda) e^{-\tau\left(\lambda\right)}
\label{eq:cf00_dust_a}
\end{equation}
with
\begin{equation}
\tau(\lambda) = \tau_V \left(\frac{\lambda}{5500\,\angstrom}\right)^{-1.3},
\label{eq:cf00_dust_b}
\end{equation}
where $L_\mathrm{ext}(\lambda)$ and $L(\lambda)$ are the attenuated and unattenuated luminosities respectively, $\lambda$ is the rest-frame wavelength of the emission line, and $\tau_V$ is the V-band (5500\AA) optical depth.
Thus the absorption curve is described by only one parameter, $\tau_V$.

The radial variation of the dust content of galaxies is not well known.
For simplicity we shall therefore assume the optical depth to be constant across the whole galaxy.
We discuss the appropriateness of this assumption in Section~\ref{sec:application_results_dust}.

It should be noted that, even aside from the lack of radial variation, this dust model is relatively basic.
We have assumed the galaxy to be infinitesimally thin, and we do not include any radiative transfer effects along the line-of-sight.
Approximating the galaxy in this way as a thin disc becomes highly questionable for highly-inclined ($\ga70\degr{}$) galaxies and we do not claim that our model works for such edge-on systems.

\subsubsection{Summary}

We have now outlined how we assign the emission-line luminosities.
All told there are five free parameters: the total star formation rate of the galaxy, $\textrm{SFR}_\mathrm{tot}$, the central metallicity, $\log_{10}Z_0$, the metallicity gradient, $\nabla_r\left(\log_{10}Z\right)$, the ionization parameter at solar abundance, $\log_{10}U_{\sun}$, and the V-band optical depth, $\tau_V$.
In the next section we discuss the fitting of our model, and the bounds we place on these parameters.

As a final cautionary note we highlight that the model only describes the nebular emission from star-forming regions.
In the centres of galaxies, however, active galactic nuclei (AGN) and low-ionization nuclear emission-line regions (LINERs) can contribute significantly to the emission-line flux.
Therefore this model should not be applied to galaxies that present signs of significant AGN/LINER contamination.

\subsection{Model fitting}

In the preceding sections we have described our model which we will use to derive the metallicity of barely resolved galaxies.
Of the modelled parameters the most scientifically interesting are the central metallicity, $\log_{10}Z_0$, and the metallicity gradient, $\nabla_r\left(\log_{10}Z\right)$.
We would like to derive meaningful errors, accounting for the degeneracies among the parameters.
Such a problem naturally lends itself to a Markov chain Monte Carlo (MCMC) approach.
Here we use the \textsc{MultiNest} algorithm \citep{2009MNRAS.398.1601F,2008MNRAS.384..449F,2013arXiv1306.2144F} accessed through a \textsc{Python} wrapper \citep{2014A&A...564A.125B}.
In light of the known degeneracies between metallicity and ionization-parameter we anticipate that the likelihood surface may be similarly degenerate.
For this reason we have adopted the \textsc{MultiNest} algorithm, which is efficient at sampling multimodal and/or degenerate posterior distributions.

\subsubsection{Prior probability distributions (Priors)}\label{sec:priors}

For the Bayesian computation we place an initial probability distribution (prior) on each parameter.
We set the priors to be all independent of one another, described as follows:

\begin{itemize}

\item $\textrm{SFR}_\mathrm{tot}$: The total SFR of the galaxy provides the overall flux normalization of the model, we place a flat prior on the interval $\left[0, 100\right]$\,$\textrm{M}_{\sun}\,\mathrm{yr}^{-1}$.
This sufficiently covers the expected range of galaxies we could observe.

It may seem more logical to adopt a logarithmic prior for this normalization constant.
Adopting such a prior caused our model to converge to local minima in our highest S/N tests (\S\ref{sec:snr_tests}).
Real data, which has much lower S/N, will not suffer the same convergence issues as the likelihood surface will be smoother.
For consistency we adopt a uniform prior throughout this paper.
This does not affect our conclusions.

\item $\log_{10}Z_0$: We place a flat prior on the central metallicity, $\log_{10}Z_0$, (logarithmic over $Z_0$).
The interval is chosen to match the full metallicity range allowed by the photoionization-model grid (\mbox{$\sim$[-1.30,0.70]}\,dex).

\item $\nabla_r\left(\log_{10}Z\right)$: We set a flat prior on the metallicity gradient of galaxies spanning the range $\left[-0.5, 0.5\right]\,\textrm{dex}/\textrm{kpc}$.
Current evidence suggests galaxies at high redshifts ($z \ga 1$) may exhibit metallicity gradients steeper than those found in lower redshift galaxies.
Typically high redshift galaxies have metallicity gradients between $-0.1$ and $0.1\,\textrm{dex}/\textrm{kpc}$, and at most $-0.3\,\textrm{dex}/\textrm{kpc}$ \citep{2016ApJ...820...84L}.
Our prior is therefore sufficiently broad to incorporate even the steepest gradients.

It should be noted that a flat prior on a metallicity gradient is not an uninformative prior.
A uniform prior in gradient is \emph{not} uniform in angle, but is biased towards steeper profiles \citep[see][]{2014arXiv1411.5018V}.
Furthermore, a minimally informative prior would yield equal probability to find any metallicity at all radii, $r$.
I.e. the 2D ($r$, $\log_{10}Z_0$) space should be evenly sampled.
Since we clip our metallicities to a finite grid of photoionization models this is difficult to achieve perfectly.
Therefore, for the simplicity of this paper we adopt a uniform prior on the metallicity gradient.
The choice of this prior will have to be revisited in future work.
We further discuss the effect of this prior in Appendix~\ref{sec:model_systematics}.

\item $\log_{10}U_{\sun}$: The photoionization-model grid already sets bounds on the allowed values of $\log_{10}U$.
We set a flat prior on $\log_{10}U_{\sun}$ such that $\log_{10}U$ can span this full range, at any metallicity.
For this paper this range is $\sim$[-5.02,-1.42]\,dex.
Remember that ultimately $\log_{10}U\left(r\right)$ will clipped to remain within the photoionization-model grid.

\item $\tau_V$: We place a flat prior on the V-band optical depth on the interval $\left[0, 4\right]$.
This should be sufficient to include all galaxies we are interested in, which have relatively strong emission-lines.

\end{itemize}

\subsubsection{Likelihood function}

The likelihood function assigns the probability that, for a given model, we would have measured the observed emission-line fluxes.

We will have a set of observed fluxes, $F_{\mathrm{obs},i}$, for each observed emission-line and for each spatial bin.
Correspondingly we have a set of errors, $\sigma_{\mathrm{obs},i}$, estimated from the data.
Our model predicts a complementary set of fluxes, $F_{\mathrm{model},i}$.
Following \citet{2004MNRAS.351.1151B}, we additionally assign a constant 4\% theoretical error, $\sigma_{\mathrm{model},i} = 0.04\,F_{\mathrm{model},i}$.

We assume that the observed fluxes, $F_{\mathrm{obs},i}$, are related to the true fluxes, $F_{\mathrm{true},i}$, through
\begin{equation}
F_{\mathrm{obs}_i} = F_{\mathrm{true}_i} + \epsilon_i,
\end{equation}
where the noise, $\epsilon_i$, is drawn from a Student's t-distribution.
Our likelihood function is therefore
\begin{equation}
\mathcal{L}(x_1,\ldots,x_n\mid\nu,\sigma_1,\ldots,\sigma_n) = \prod_{i=1}^{n} \mathcal{L}(x_i\mid\nu,\sigma_i)
\end{equation}
with
\begin{equation}
\mathcal{L}(x_i\mid\nu,\sigma_i) = \frac {\Gamma\left(\frac{\nu+1}{2}\right)} {\Gamma\left(\frac{\nu}{2}\right)\sqrt{\pi\nu}\sigma_i} \left(1 + \frac{1}{\nu}\left(\frac{x_i}{\sigma_i}\right)^2\right)^{-\frac{\nu+1}{2}},
\end{equation}
where we define the residual as
\begin{equation}
x_i = F_{\mathrm{obs},i} - F_{\mathrm{model},i},
\end{equation}
and the square of the scale parameter as
\begin{equation}
\sigma_i^2 = \frac{\nu - 2}{\nu}(\sigma_{\mathrm{obs},i}^2 + \sigma_{\mathrm{model},i}^2).
\end{equation}
In this paper we assume $\nu = 3$ degrees of freedom.

There are two motivations for adopting Student's t-distribution over the more traditional normal distribution.
The first and highly practical reason is to add robustness to our fitting.
Student's t-distribution is more heavily tailed than the normal distribution.
Therefore outliers with large residuals will be penalized less by Student's t-distribution than by the normal distribution.
Even if most of the data is well described by the normal distribution, one errant data point can have disastrous consequences on the inference.
Essentially by adopting a more robust likelihood function we are trading an increase in accuracy for a decrease in precision.

The second reason for adopting Student's t-distribution is that in fact our data may indeed be better described by Student's t-distribution than the normal distribution.
The emission-line fluxes are typically measured from spectra where the resolution is such that the emission line is covered only by a few wavelength elements.
In this case the associated errors are calculated only from a few independent pieces of information, and hence the Student's t-distribution is more appropriate.
Precisely calculating the degrees of freedom of each emission-line is difficult, although in theory can be estimated from repeat observations.
For simplicity we assume the number of degrees of freedom is small, and hence we choose a constant $\nu = 3$ degrees of freedom.

\subsection{PSF model}\label{sec:psf_model}

There is one further aspect of the model that we have not yet discussed.
The galaxy model fluxes are distributed assuming a PSF.
To derive meaningful results from the best fit model it is important to input a PSF that closely matches the true seeing of the observations.
The adopted PSF should therefore be driven by the data itself.

In this paper we will use MUSE observations of the Hubble Deep Field South \citep{2015A&A...575A..75B}.
The authors use a moderately bright star also within the MUSE field of view (FoV) to derive the PSF.
The best-fit Moffat profile for this star has the parameters as given in Table~\ref{tab:psf_params}.
For consistency, unless otherwise specified, we will adopt this empirical model throughout this paper as our fiducial PSF.

\begin{table}
\caption{Moffat parameters of the adopted PSF model, indicating knots of a piecewise-linear interpolation.
Each wavelength has an associated full-width half-maximum size (FWHM) and a Moffat--$\beta$ parameter.}
\label{tab:psf_params}
\begin{tabular}{ccc}
\hline
Wavelength & FWHM & $\beta$ \\
$[\angstrom{}]$ & $[\textrm{arcsec}]$ &  \\
\hline
4750 & 0.76 & 2.6 \\
7000 & 0.66 & 2.6 \\
9300 & 0.61 & 2.6 \\
\hline
\end{tabular}
\end{table}

\section{Model Testing}\label{sec:model_testing}

In the previous section we presented our method for modelling the emission lines of distant galaxies.
Before moving to the modeling of distant galaxies in the following section, we here assess the reliability of our model.
Of all the modelled quantities, we are most interested in the metallicity profile, hence we will only focus on validating two of the model's parameters: the central metallicity, and the metallicity gradient.
In essence we consider $\textrm{SFR}_\mathrm{tot}$, $\log_{10}U_{\sun}$ and $\tau_V$ all to be nuisance parameters.

Here we present two categories of tests.
In the first set of tests (\S\ref{sec:accu_prec_tests}) we fit the model to mock data constructed using noisy realizations of the model itself.
This will allow us the observe intrinsic systematics and uncover inherent limitations of our method.
However, these tests cannot assess whether our model is actually a good description of a real galaxy.
So, to answer this we present a second set of tests (\S\ref{sec:real_tests}) using mock data from downgraded observations of low redshift galaxies.
With these we can study how the model performs for realistic galaxies with complex structure, violating our idealized model assumptions.

\subsection{Accuracy and precision tests}\label{sec:accu_prec_tests}

In order to validate our method we must minimally show that the model can recover itself.
With the inclusion of noise it is not obvious that this should be the case.
A combination of low S/N and resolution loss may yield highly degenerate model solutions.

In the following tests we use our model to construct simulated mock observations for a galaxy at a redshift of $z=0.5$, using the PSF given in Table~\ref{tab:psf_params}.
We assume the star forming disc of the galaxy to have an exponentially declining star-formation rate density 
\begin{equation}
\label{eq:exp_disc}
\Sigma_\textrm{SFR} \propto e^{-r/r_d}
\end{equation}
where $r_d$ is the exponential scale-length of the disc.
With our model we generate four noise-free emission-line images\footnote{\forbidden{O}{ii}{3726,3729}, \Hgamma{}, \Hbeta{}, and \forbidden{O}{iii}{5007}}.
To this data we add normally distributed noise, with the standard deviation depending on the pixel flux $F_i$ as follows
\begin{equation}
\label{eq:snr}
\sigma_i = \alpha \sqrt{F_i},
\end{equation}
where $\alpha$ is a scaling factor.
This scaling factor is the same for all emission lines.
By adjusting the scaling factor we can achieve different S/N observations.
We define the S/N as that of the brightest pixel in the unbinned \Hbeta{} map.

We must treat the fake data as we would for real data, therefore we bin spaxels together to reach a minimum $\textrm{S/N} = 5$ in all emission lines.
This binning algorithm is outlined in Appendix~\ref{sec:binning_algorithm}.

\subsubsection{Varying S/N}\label{sec:snr_tests}

Our solution should converge to the true solution at high S/N, but might be biased or show incorrect uncertainty estimates at lower S/N.
In the following we therefore explore a range of S/N levels ($\textrm{S/N} = 3,6,9,50$).

For the test we construct 50 realisations of mock data, at a given S/N ratio.
For each realisation we fit the model and retrieve marginal posterior probability distributions of the two parameters of interest (the central metallicity, $\log_{10}Z_0$, and metallicity gradient, $\nabla_r\left(\log_{10}Z\right)$).
We take the median of each marginal posterior to be the best-fit solution.

\begin{figure}
\includegraphics[width=\linewidth]{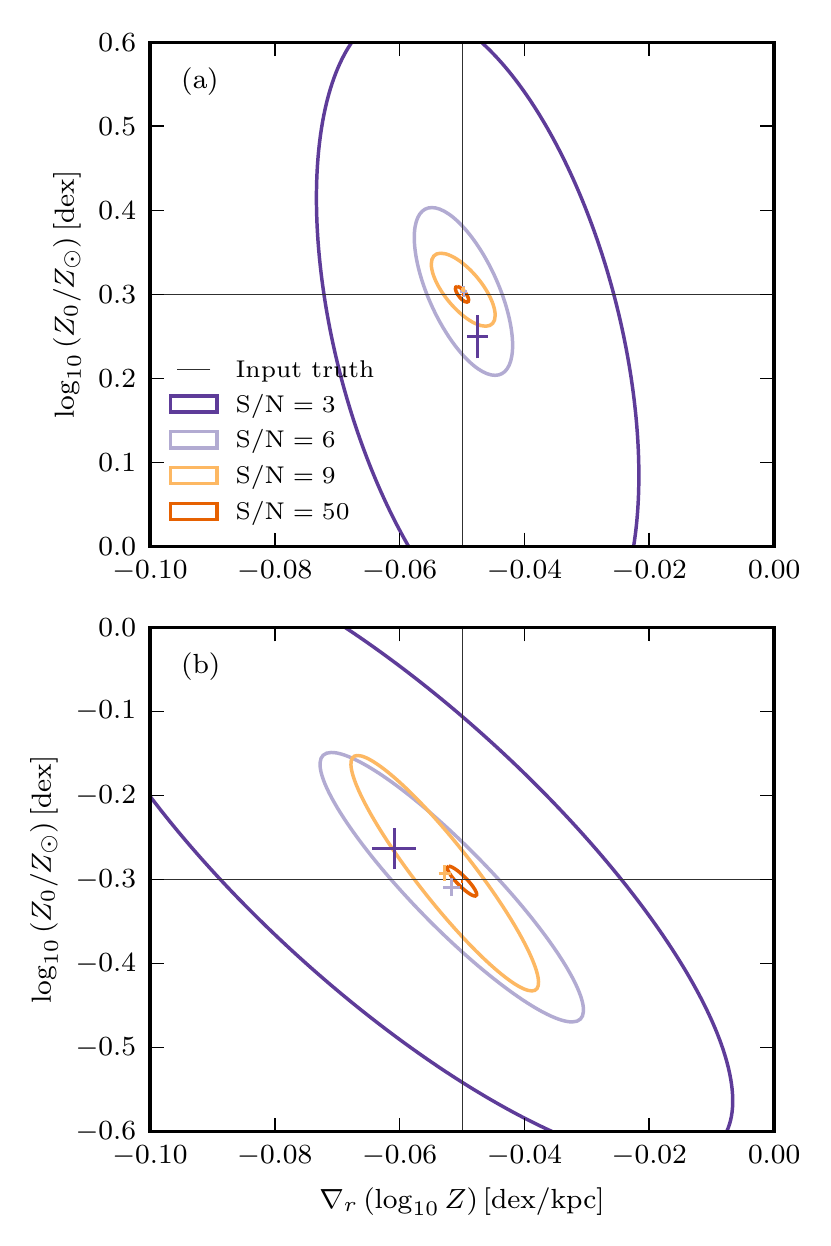}
\caption{
The effects of S/N on accuracy and precision of the inferred central metallicity, $\log_{10}Z_0$, and metallicity gradient, $\nabla_r\left(\log_{10}Z\right)$.
Plot showing error ellipses for varying S/N, drawn such that they enclose 90\% of the scatter (assuming the data to be distributed normally).
Coloured error crosses indicated the means (and standard error on the mean) at each S/N level.
The two different panels show this experiment for two different sets of original model inputs.
In panel (a) Model inputs were $\log_{10}\left(Z_0 / Z_{\sun}\right) = 0.3\,\textrm{dex}$, $\nabla_r\left(\log_{10}Z\right) = -0.05\,\textrm{dex}/\textrm{kpc}$, $\textrm{SFR}_\mathrm{tot} = 1\,\textrm{M}_{\sun}\,\mathrm{yr}^{-1}$, $r_d = 0.4\arcsec$,  $\log_{10}U_{\sun} = -3\,\textrm{dex}$, $\tau_V = 0.7$.
In panel (b) Model inputs identical to (a) except for $\log_{10}\left(Z_0 / Z_{\sun}\right) = -0.3\,\textrm{dex}$.
}
\label{fig:snr_tests}
\end{figure}

In Fig.~\ref{fig:snr_tests} we show the mean and scatter of these best-fit values over the 50 realizations.
We provide this for a range in S/N levels, and for two slightly different input models (Panels a \& b).
From this we can assess that at all but the lowest S/N level there is little systematic offset of the mean from true value.
For $\textrm{S/N} \geq 6$ we find that bias on the central metallicity is $\mbox{<0.01}\,\textrm{dex}$ and on the metallicity gradient $\mbox{<0.003}\,\textrm{dex}/\textrm{kpc}$.
At $\textrm{S/N} = 3$ there is some noticeable offset, but the realization-to-realization scatter is much larger.
We discuss biases in more detail in Appendix~\ref{sec:model_systematics}.
Therein we explore a larger portion of the parameter space where strong systematic offsets can arise.

The tests here also show that there is considerable scatter in the poor S/N=3 data. 
This is of course unsurprising, however, even the good S/N=9 results in Fig.~\ref{fig:snr_tests}(b) show moderate scatter.
Since we are performing an MCMC fit, we retrieve the full posterior probability distribution (or \textit{posterior} for short).
We can use the 50 repeat realizations to infer whether the posterior is a good estimate of this error.
For each realisation we define the z-score to be the difference between the true value and the estimated mean in units of the predicted uncertainty.
If the uncertainty estimates are accurate, these z-scores should be distributed as a standard normal distribution (zero mean and unit variance).
In Tables~\ref{tab:error_test_logZ0}~\&~\ref{tab:error_test_dlogZ} we summarize these z-scores for the model shown in Fig.~\ref{fig:snr_tests}(b).
We see that the tabulated percentages are slightly smaller than would be expected.
This indicates that our posteriors typically underestimate the true error.
However, this is only a relatively small difference so, although not perfect, we conclude these error estimates to be acceptable.
For reference we also present Q-Q plots in the appendix (Fig.~\ref{fig:QQ}), comparing the z-scores to a theoretical normal distribution.

\begin{table}
\caption{Percentage of 50 repeat realizations with $\log_{10}\left(Z_0\right)$ z-scores within a given range.
Associated Q-Q plot are found in the appendix (Fig.~\ref{fig:QQ}).
Results here are for the model shown in Fig.~\ref{fig:snr_tests}(b).}
\label{tab:error_test_logZ0}
\begin{tabular}{ccccc}
\hline
S/N & $-1 \le z < 0$ & $0 \le z < 1$ & $-1 \le z < -1$ & $-2 \le z < 2$ \\
\hline
3 &  (22 $\pm$ 3)\% & (46 $\pm$ 4)\% & (68 $\pm$ 3)\% & (98 $\pm$ 1)\% \\
6 &  (28 $\pm$ 3)\% & (30 $\pm$ 3)\% & (58 $\pm$ 3)\% & (84 $\pm$ 3)\% \\
9 &  (28 $\pm$ 3)\% & (26 $\pm$ 3)\% & (54 $\pm$ 4)\% & (88 $\pm$ 2)\% \\
50 &  (30 $\pm$ 3)\% & (34 $\pm$ 3)\% & (64 $\pm$ 3)\% & (90 $\pm$ 2)\% \\
\hline
Expected & 34\% & 34\% & 68\% & 95\% \\
\hline
\end{tabular}
\end{table}

\begin{table}
\caption{Percentage of 50 repeat realizations with $\nabla_r\left(\log_{10}Z\right)$ z-scores within a given range.
Associated Q-Q plot are found in the appendix (Fig.~\ref{fig:QQ}).
Results here are for the model shown in Fig.~\ref{fig:snr_tests}(b).}
\label{tab:error_test_dlogZ}
\begin{tabular}{ccccc}
\hline
S/N & $-1 \le z < 0$ & $0 \le z < 1$ & $-1 \le z < -1$ & $-2 \le z < 2$ \\
\hline
3 &  (40 $\pm$ 3)\% & (10 $\pm$ 2)\% & (50 $\pm$ 4)\% & (84 $\pm$ 3)\% \\
6 &  (26 $\pm$ 3)\% & (32 $\pm$ 3)\% & (58 $\pm$ 3)\% & (86 $\pm$ 2)\% \\
9 &  (22 $\pm$ 3)\% & (32 $\pm$ 3)\% & (54 $\pm$ 4)\% & (90 $\pm$ 2)\% \\
50 &  (26 $\pm$ 3)\% & (28 $\pm$ 3)\% & (54 $\pm$ 4)\% & (90 $\pm$ 2)\% \\
\hline
Expected & 34\% & 34\% & 68\% & 95\% \\
\hline
\end{tabular}
\end{table}

\subsubsection{Varying PSF}

The preceding section showed that at moderate to high S/N, our model is unbiased when fitting itself.
These tests were performed with decent spatial resolution ($r_d \ga 0.5 \times \textrm{FWHM}$), so we will now explore the effect of degrading the PSF.
To do this, we create a series of mock data with fixing the physical model parameters, but with different PSFs. 

We model changes in the seeing simply through changes in the FWHM of the PSF.
The wavelength dependence of the seeing is retained, and we modulate the FWHM amplitude by a multiplicative factor.
The Moffat $\beta$ parameter remains fixed. 
We remind the reader that our S/N is defined on the peak (unbinned) flux of the \Hbeta{} emission line (\S\ref{sec:accu_prec_tests}), so by changing the PSF we inadvertently alter the S/N.
To isolate the effects of resolution from those of S/N, we shall keep $\alpha$ (the noise scaling factor in equation \ref{eq:snr}) fixed to that used for the fiducial PSF.
The total flux from the galaxy remains unchanged.

\begin{figure}
\includegraphics[width=\linewidth]{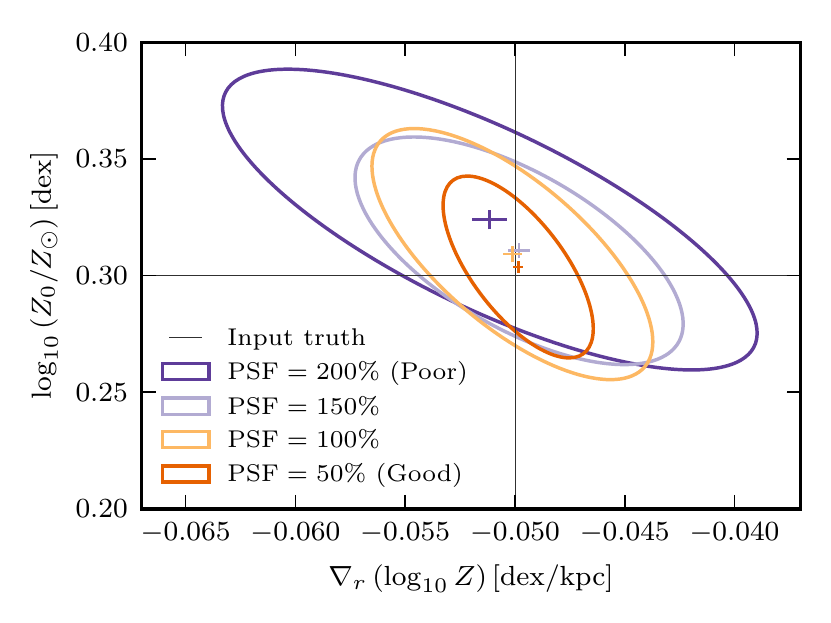}
\caption{
Effects of changing the PSF on the inferred central metallicity and metallicity gradient.
We show error ellipses for a series of improving PSFs (see Fig.~\ref{fig:snr_tests} for plot description).
Here a 200\% PSF indicates observations with a FWHM double that of the fiducial (100\%) model.
The noise scaling factor ($\alpha$ in equation~\ref{eq:snr}) is fixed such that the 100\% model has a peak $\textrm{S/N} = 9$.
We adopt the same model inputs as used Fig.~\ref{fig:snr_tests}(a).
The disc scale-length is $r_d=0.4''$.
}
\label{fig:psf_tests}
\end{figure}

In Fig.~\ref{fig:psf_tests} we show the mean and scatter of 50 realizations for four different PSFs.
This shows that even with significantly poorer seeing our model is still able to recover the true values with little systematic offset.
However, poorer seeing will introduce information loss and the precision to which we can determine the metallicity gradient is much reduced.
We caution the reader that this statement can not readily be converted into an absolute FWHM of the PSF since what is of real importance here is the relative size of the PSF to the size of the galaxy.
But as a guide for the reader, the percentages in Fig.~\ref{fig:psf_tests} correspond to PSFs between $\sim 0.4 \textrm{--} 1.5''$ FWHM, which should be compared to a galaxy that has a $r_d=0.4''$ disc scale-length (which would be typical for $3\times10^{10}\,\textrm{M}_{\sun}$ disc galaxies at $z=0.75$ \citep[e.g.][]{2014ApJ...788...28V}). 

It should be noted that the direction of the systematic offset in the poor (PSF = 200\%) seeing data is actually towards a steeper metallicity gradient, rather than towards the flat gradient that one might na\"{i}vely expect.
Since seeing is wavelength dependent its effects can be complicated, and therefore worse seeing may not automatically lead to a flatter inferred gradient.
However, it is perhaps more likely a reflection of systematics intrinsic to the modelling and/or introduced by the model priors (see Appendix~\ref{sec:model_systematics}).

\subsubsection{Varying inclination}

Altering the PSF is not the only way to reduce spatial information.
Highly inclined (edge-on) galaxies lose considerable resolution along the minor axis.
We should check that our method is able to recover the same metallicity profile for a galaxy independent of its inclination.

Again we construct a series of mock observations where the only variation is in the inclination of the galaxy.
As before, in order to remove the effects of changing S/N, we fix $\alpha$ (the noise scaling factor in equation \ref{eq:snr}) to that used for the fiducial $\textrm{inc.}=0^\circ$ model.

\begin{figure}
\includegraphics[width=\linewidth]{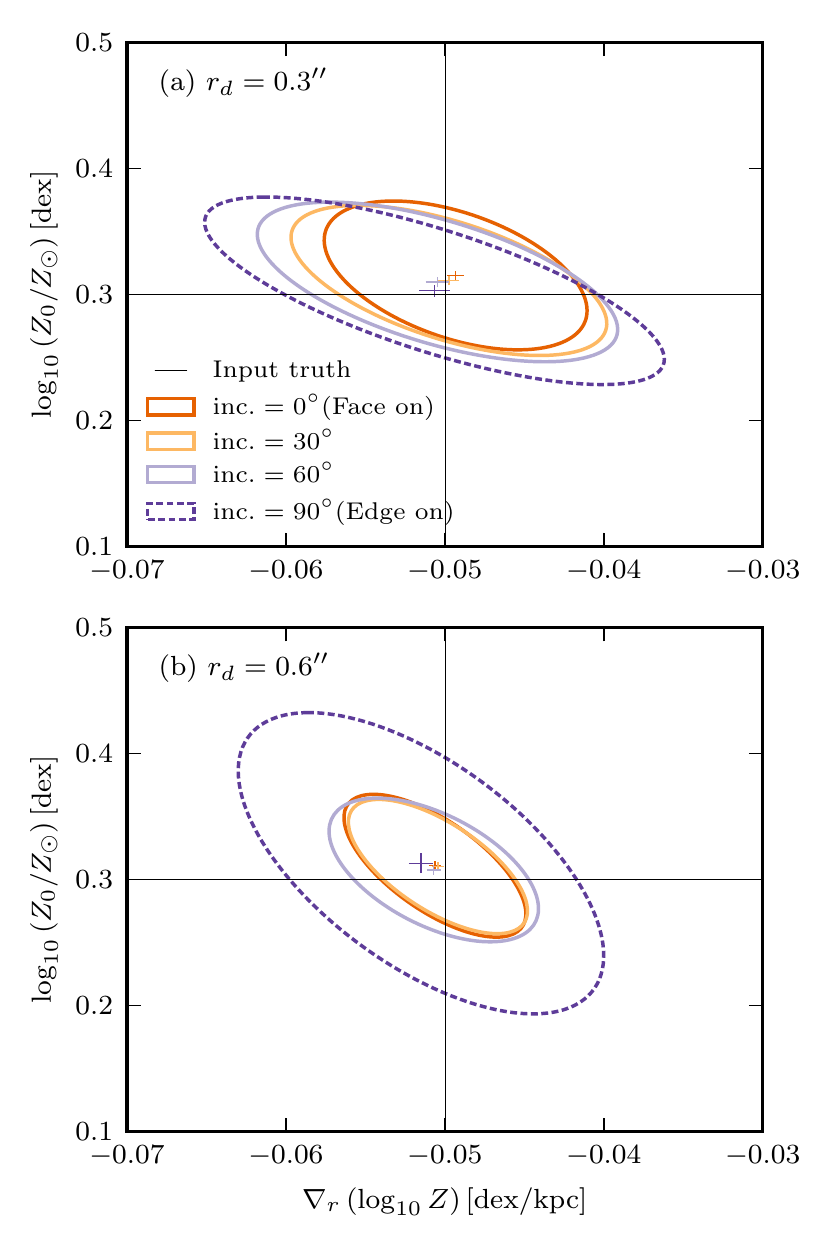}
\caption{
The impact of inclination on the accuracy and precision to which we can derive the central metallicity and metallicity gradient.
We show error ellipses for a set of progressively more inclined models (see Fig.~\ref{fig:snr_tests} for plot description).
The noise scaling factor ($\alpha$ in equation~\ref{eq:snr}) is fixed such that the $\textrm{inc.}=0^\circ$ model has a peak $\textrm{S/N} = 9$.
}
\label{fig:inc_tests}
\end{figure}

In Fig.~\ref{fig:inc_tests} we show the mean and scatter of 50 realizations for four different inclinations.
We perform this exercise for two galaxies of different sizes ($r_d=0.3\arcsec$ and  $r_d=0.6\arcsec$), where the smaller galaxy should be more sensitive to inclination effects.
It can be seen that even in the edge-on case we are able to well recover the metallicity profile, although admittedly to a lower precision than for the face-on galaxy.

It should be stressed, however, that even though the method works for the extreme edge-on cases there are significant limitations in the galaxy model at high inclinations.
Because we assume the galaxy to be infinitesimally thin, two issues arise.
Firstly, at high inclinations the centres of dusty galaxies may be obscured, but since we do not include any radiative transfer effects along the line-sight the model does not reproduce this.
Secondly, when a galaxy is nearly edge-on it becomes almost impossible to distinguish metallicity that varies with radius from metallicity that varies with vertical disc height.
Even with high-spatial resolution observations these problems would remain.
For these reasons we caution the reader that the results for highly inclined galaxies are unlikely to be relevant for real galaxies and we will limit our studies to galaxies with inclinations less than $\sim 70\degr$.

The tests presented so far are not sufficient to validate our model, and indeed further tests are required.
In the following section we use mock observations constructed from real observations of low redshift galaxies.
This will enable us to compare our model against data that more closely resembles real, rather than idealized, galaxies.

\subsection{Model tests with realistic data}\label{sec:real_tests}

So far we have ascertained that our method is able to recover the true metallicity profile.
Although adverse conditions (low S/N and poor seeing) reduce the precision of the method, they do not significantly impact upon the accuracy.
This does not, however, verify that the model is a good description of real galaxies.
To address this we will fit the model to mock data generated from observations of low redshift galaxies, downgraded in both S/N and resolution.

The mock data is constructed from IFS observations of three low redshift galaxies (UGC463, NGC628, NGC4980).
These galaxy were not selected especially to be representative of higher redshift galaxies (although their SFRs are comparable to those we will study).
Instead these galaxies were chosen primarily owing to the availability of high quality IFS data, and because they are not highly inclined galaxies.
Two of these galaxies were observed with MUSE (UGC463 and NGC4980) and the other (NGC628) was observed as part of the PPAK IFS Nearby Galaxies Survey \citep{2011MNRAS.410..313S}.
We construct emission-line maps\footnote{The exact details of how these maps are obtained are not crucial to our analysis. For a self-consistent analysis we simply require realistic mock inputs, ideally with high S/N and good spatial resolution.} of \Hbeta{}, \forbidden{O}{iii}{5007}, \Halpha{}, \forbidden{N}{ii}{6584} and \forbidden{S}{ii}{6717,6731} from these observations and convolve these maps with the seeing and bin them to the appropriate pixel scale to produce mock images.
Finally noise is added and the data binned as described above (Section~\ref{sec:accu_prec_tests}).
In the following we define the size of the galaxies using the disc scale-length of dust-corrected \Halpha{} flux profile.
Note that the galaxy centres are defined using the stellar light \textit{not} the nebular emission (which can be clumpy and asymmetric).

In addition to the emission-line images, our method requires a SFR map for each galaxy.
Typically these SFR maps will be created from high-resolution observations.
So, we generate SFR maps using the dust-corrected \Halpha{} maps of the low redshift galaxies.
These maps are then degraded to a resolution comparable to that of the Hubble Space Telescope (HST), i.e. a Gaussian PSF with $\textrm{FWHM}=0.1''$ and pixel scale $0.05''$.
We do not add any additional noise to the SFR maps.

To test our ability to measure the metallicity profile of these mock observations, we run our full model fitting procedure on galaxies of two different sizes ($r_d=0.4''$ and $r_d=0.8''$), simulated with $\textrm{S/N} = 9$, at a redshift $z=0.255$\footnote{At this redshift all five emission lines are within the MUSE wavelength coverage. More typically, however, we will apply this model to higher redshift galaxies where \forbidden{O}{ii}{3726,3729} is available, but \Halpha{}, \forbidden{N}{ii}{} and \forbidden{S}{ii}{} are not.}, and with the PSF given in Table~\ref{tab:psf_params}.
At this redshift \Hbeta{}, the most blueward emission line, is the most affected by seeing and has a $\textrm{FWHM}=0.7''$.
These results are then compared to the metallicity derived from the high-resolution (non-degraded) data. 
We compute the latter using the \textsc{IZI} procedure developed by \citet{2015ApJ...798...99B}, which solves for metallicity, marginalized over the ionization parameter.
For consistency with our galaxy model we use the same \citetalias{2013ApJS..208...10D} ($\kappa=\infty$) photoionization model grid.
We fit a simple exponential model for the metallicity as a function of radius (i.e. equation~\ref{eq:metallicity_profile}), where each data point is weighted proportional to its \Halpha{} flux.
We weight by flux because unless one can resolve \ion{H}{ii} regions individually, one is unavoidably weighted towards the emission-line ratios of the brightest \ion{H}{ii} regions. Thus, for comparison to our low-resolution mock data, it is appropriate to weight our fit by the \Halpha{} flux.
We caution the reader that the high-resolution metallicity profiles presented here should not be considered definitive.
The analysis that follows is nonetheless self-consistent.

\begin{figure*}
\includegraphics[width=\linewidth]{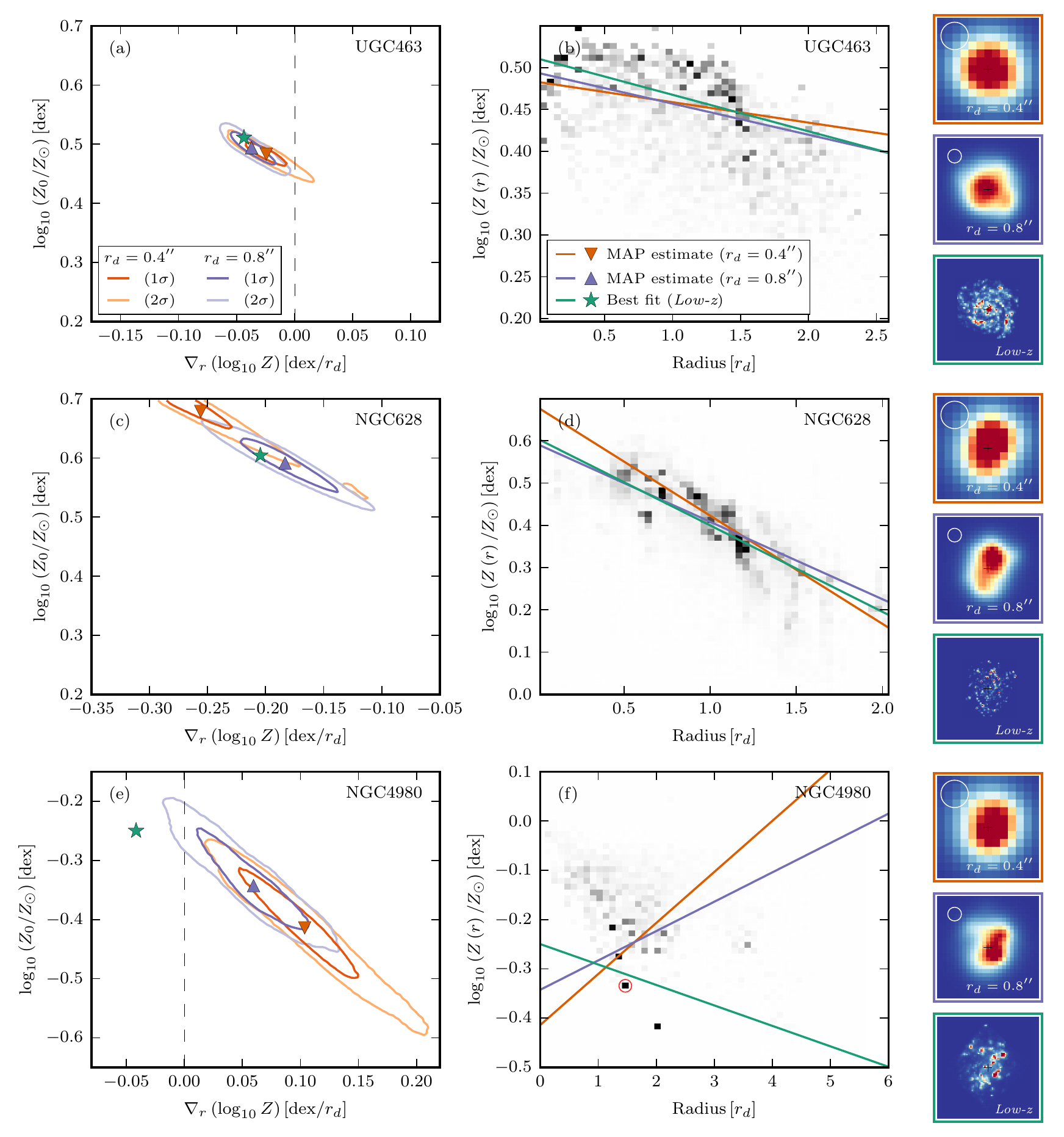}
\caption{Comparison between the true and model derived metallicity profiles for three galaxies: UGC463, NGC628 and NGC4980, shown in descending order.
(Left) We show the marginalized 2D probability contours for the central metallicity, $\log_{10}Z_0$, and metallicity gradient, $\nabla_r\left(\log_{10}Z\right)$ (after stacking 50 mock realizations).
Results are shown for two mock galaxies of different sizes: $r_d=0.4''$ (orange) and $r_d=0.8''$ (blue).
In addition to the $1\sigma$ \& $2\sigma$ contours, we plot the MAP estimates as triangles.
N.B. panels (a,c,e) are all scaled to span the same axis ranges.
(Centre) Using the full resolution data we construct a 2D histogram of metallicity versus radius.
We weight the histogram by the \Halpha{} flux of each data point.
Overploted are the MAP solutions for the $r_d=0.4''$ and $r_d=0.8''$ models (orange and blue respectively).
Additionally we also show the exponential best-fit to the full resolution data (green).
The locations of the the best-fit parameters for the full resolution data are indicated on the left as a green star.
Histograms are plotted on a linear scale, clipped between the 1\textsuperscript{st} and 99\textsuperscript{th} percentiles.
In panel (f) we indicate one bin with a red circle. 
This single bin contains 10\% of the total \Halpha{} flux.
(Right) We show aligned images of the \Hbeta{} emission line for the two mocks and the full resolution data.
The images are shown without noise, and are plotted on a linear scale, clipped between the 1\textsuperscript{st} and 99\textsuperscript{th} percentiles.
The white circle indicates a $0.7''$ FWHM PSF in the mock images.}
\label{fig:real_tests}
\end{figure*}

In Fig.~\ref{fig:real_tests} we present a comparison of the inferred and true metallicity profiles.
For each mock dataset we create 50 realizations and calculate the marginalized 2D probability on the central metallicity, $\log_{10}Z_0$, and metallicity gradient, $\nabla_r\left(\log_{10}Z\right)$.
The left-hand panels show this marginalized probability, after stacking all 50 realizations.
A triangle indicates the maximum a posteriori (MAP) estimate of this stacked marginalized probability.
In the central panels we present the true metallicity profile, with the best-fit exponential model and MAP estimate models overplotted.
As can be seen, our model performs well for UGC463 and NGC628, but derives an entirely different solution for NGC4980.
We shall now discuss each galaxy in turn.

\begin{description}
\item[\textit{UGC463}] This is a SAB(rs)c galaxy \citep[][herein V91]{1991rc3..book.....D}\defcitealias{1991rc3..book.....D}{V91} and has a stellar mass $\log_{10}\left(\textrm{M}_\ast/\textrm{M}_{\sun{}}\right) = 10.6$ \citep{2013A&A...557A.131M}.
This galaxy was observed during MUSE commissioning (Martinsson et al. in prep.).
Before we downgrade them, the physical resolution of the observations is $\sim 240\,\textrm{pc}$.
The convolved images indicate that the galaxy is roughly axisymmetric, with the brightest flux consistent with the centre of the galaxy.
From panel (a) we note that both the inferred model solutions are in agreement with the best fit to the high-resolution data.
Despite the $r_d=0.4''$ MAP metallicity gradient estimate being a factor two shallower than the best fit, panel (b) shows this solution is still consistent with the data.
In fact it could be argued that no solution is an exceptionally good description of the data.
The data indicates the galaxy has a downturn in metallicity beyond $r\ga1.3\,r_d$ and therefore does not support any simple exponential metallicity profile.

We actually find it quite unexpected that the model succeeds in recovering the metallicity profile.
This is because the galaxy demonstrably breaks our assumption that the ionization parameter is anti-correlated to the metallicity (equation~\ref{eq:ionization_parameter_coupling}).
In this galaxy the ionization parameter and metallicity are in fact positively correlated (see Fig.~\ref{fig:UGC463_logZlogU}).
Nevertheless the model is perfectly able to recover the truth, although since this is a single case it is not possible generalise about the robustness of our model.
We can, however, infer that our derived metallicity gradients are not entirely driven by ionization parameter gradients in galaxies.

\item[\textit{NGC628}] This galaxy, like the previous, appears to be a SA(s)c galaxy \citepalias{1991rc3..book.....D} with stellar mass $\log_{10}\left(\textrm{M}_\ast/\textrm{M}_{\sun{}}\right) = 10.3$ \citep{2015ApJS..219....5Q}.
Before we downgrade it, the galaxy physical resolution of the data is $\sim 120\,\textrm{pc}$.
Dissimilarly, however, NGC628 has a dearth of star forming regions in its centre.
This is accentuated by the $r_d=0.8''$ image the galaxy, which is visibly lopsided and features a strong star forming complex to the upper-right of the centre.
Panel (c) indicates that in the $r_d=0.8''$ case our model is able to recover the same result as the best fit.
Whereas for the smaller $r_d=0.4''$ case the model appears to perform less well, and is mildly inconsistent with the best fit solution.
Notably the solution for the $r_d=0.4''$ case favours a steeper metallicity profile than $r_d=0.8$ solution.
It is interesting to note that in this case, with significant emission line flux outside the central region, worse seeing does not lead automatically to a shallower metallicity gradient, which one might na{\"\i}vely expect.

On examination of panel (d), however, it becomes clear that the $r_d=0.4''$ MAP estimate is not actually a bad description of the data and arguably provides a better characterization of the data than either the $r_d=0.8''$ MAP estimate or high-resolution best fit.
A plausible explanation is that with worsening resolution, we become increasingly weighted towards the metallicity of the brightest \ion{H}{ii} regions.
In the high-resolution case it appears that the metallicity trend deviates from linear in this galaxy, and the small scale structure of the metallicity profile plays a central role.
When the relative importance of the PSF is larger (i.e. in the $r_d=0.4''$ case) these features are smeared out and the fit is no longer affected by these structures.
It should be noted that even supplying a very high resolution SFR map does not resolve this issue.
A combination of the seeing and finite S/N produces an irreversible loss of information.

We direct the interested reader towards a similar study by \citet{2014A&A...561A.129M} who also study resolution effects on the metallicity gradient with NGC628 amongst other galaxies.

\item[\textit{NGC4980}] This galaxy was observed as part of the MUSE Atlas of Disks (MAD) (Carollo et al. in prep.).
It is a SAB(rs)a pec? galaxy \citepalias{1991rc3..book.....D} and has a stellar mass $\log_{10}\left(\textrm{M}_\ast/\textrm{M}_{\sun{}}\right) = 9.2$ \citep{2015ApJS..219....5Q}.
Before downgrading, the physical resolution of the data is $\sim 80\,\textrm{pc}$.
Spiral structure is not readily evident in the \Hbeta{} images, instead the emission-line flux is dominated by a few \ion{H}{ii} regions.
NGC4980 is extremely clumpy, for example $\sim10\%$ of the total \Halpha{} flux is contained within one spaxel.
As shown in panel (e), both the $r_d=0.4''$ and $r_d=0.8''$ MAP solutions are consistent with one another.
However, they are both inconsistent with the best fit solution to the extent that they even have the opposite sign for the metallicity gradient.

Panel (f) shows the true metallicity profile of the galaxy.
The lower surface brightness emission supports a flat or slightly negative metallicity gradient.
But the flux is dominated by a few bright \ion{H}{ii} regions which have metallicities significantly lower than fainter \ion{H}{ii} regions at the same radius.
As a result, none of the solutions (including the low-z best fit) provide a good depiction of the data.
It should be stressed that the model parameter uncertainties estimate the impact of the random data errors, however, by definition they do not account for the systematic errors caused by applying the wrong model.

It is challenging to define a meaningful metallicity gradient in galaxies like NGC4980.
At low redshift one could potentially treat the bright low-metallicity \ion{H}{ii} regions as outliers from the true metallicity profile.
Whereas as at higher redshifts one would treat the brightest emission as representative of the metallicity profile.
\end{description}

\begin{figure*}
\includegraphics[width=\linewidth]{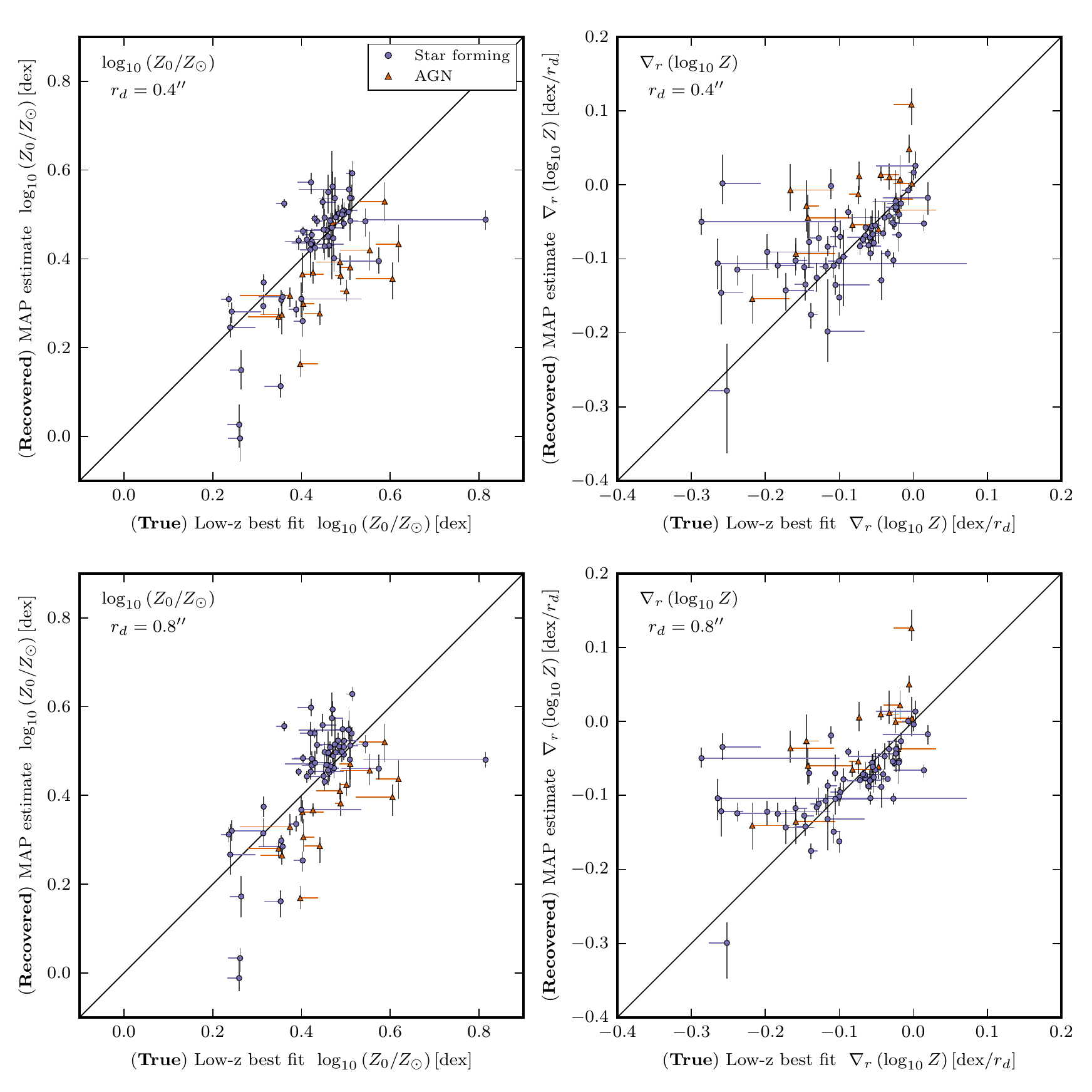}
\caption{Assessment of the models ability to recover the ``true'' metallicity profile for a sample of 76 CALIFA galaxies.
As before, we simulate mock versions of each galaxy at 2 different sizes, $r_d=0.4''$ (top) and $r_d=0.8''$ (bottom).
(Left) We plot the model derived value for the central metallicity vs the true value derived from the undegraded data.
(Right) Similarly, we compare the model derived metallicity gradient.
In each panel galaxies are represented by blue circles or orange triangles, the former indicating regular star-forming galaxies and the latter indicating galaxies with AGN.
The vertical errorbars indicated the $1\sigma$ errors reported by the model fit.
The horizontal ``errorbars'' do \emph{not} indicate the statistical error in the true gradient, but rather they indicate by how much the result would change if the true profile was instead determined from azimuthally averaged data, see text for details.
We indicate the 1:1 relation with a black line.
If our model is good at recovering the true metallicity profile we would expect most galaxies should lie along this line.
}
\label{fig:real_tests_summary}
\end{figure*}

Testing our model against these three galaxies has shown that our method does indeed have the power to recover the metallicity profile even at the marginally resolved limit.
However, for one of the galaxies our model fails catastrophically.
Clearly a larger sample is required to assess whether such cases are common.

We repeat the previous exercise, downgrading IFS observations with a larger sample of nearby galaxies selected from the 3\textsuperscript{rd} CALIFA Data Release \citep{2016A&A...594A..36S, 2012A&A...538A...8S, 2014A&A...569A...1W}.
From this we select a sub-sample that has morphological information (RA, Dec., inc., PA) provided by HyperLEDA \citep{2014A&A...570A..13M}.
We exclude galaxies that are either highly-inclined ($\ge 70\degr{}$), have low \Halpha{} SFR ($<1\,\textrm{M}_{\sun{}}\,\textrm{yr}^{-1}$), or are very small ($r_d < 7''$).
After pruning the sample, 76 CALIFA galaxies remain.
For each of these galaxies we downgrade images of their emission lines\footnote{\Hbeta{}, \forbidden{O}{iii}{5007}, \Halpha{}, \forbidden{N}{ii}{6584} and \forbidden{S}{ii}{6717}} and use our model to recover the metallicity profile.

In Fig.~\ref{fig:real_tests_summary} we compare the model recovered values of the central metallicity ($\log_{10}Z_0$) and the metallicity gradient ($\nabla_r\left(\log_{10}Z\right)$) against those derived from the full-resolution data.
For this we employ two methods of determining the true metallicity profile in the full-resolution data.
Our primary method is the same as before, where we perform a \Halpha{} flux weighted linear-fit to the metallicity derived in the individual CALIFA spaxels.
The metallicity is computed using \textsc{izi} in the spaxels that have all emission lines (\forbidden{O}{ii}{3726,3729}, \Hbeta{}, \forbidden{O}{iii}{5007}, \Halpha{}, \forbidden{N}{ii}{6584} and \forbidden{S}{ii}{6717,6731}) with $\textrm{S/N} > 3$.
We exclude spaxels that do not have \forbidden{O}{iii}{}/\Hbeta{} and \forbidden{N}{ii}{}/\Halpha{} line-ratios consistent with emission from star-formation.
Unfortunately individual spaxels may not have sufficient S/N which could bias our metallicity profile towards that of the brightest \ion{H}{ii} regions.
Therefore to assess the impact this might have we employ a second method for determining the true metallicity profile.
Instead of using individual spaxels, we first integrate the flux into elliptical annuli (with major width $4''$) before deriving the metallicity in each.
This avoids excluding low-luminosity \ion{H}{ii} regions that, while faint, could be numerous enough have a non-negligible contribution to the total flux.
This second method is somewhat limited, however, and might be skewed by the emission of diffuse ionized gas particularly in the outskirts of the galaxies.
With this caution in mind, we indicate both results in Fig.~\ref{fig:real_tests_summary}, where the data points represent the fit to individual spaxels, and the end of the horizontal ``errorbar'' is situated at the location of the fit to the annularly binned data.
It can clearly be seen that for most galaxies there is little difference between the binned and unbinned methods.
However, a few galaxies do show large differences, indicating that a ``true'' metallicity profile for these galaxies is perhaps poorly defined.

In the figure we observe that there is a good agreement between the results recovered by the model and the low-z best fit, with most galaxies lying close to the 1:1 line.
Many of the galaxies that lie off the 1:1 line possess AGN (shown as triangles in the plot).
We define galaxies as possessing an AGN if the innermost annular bin has \forbidden{O}{iii}{}/\Hbeta{} and \forbidden{N}{ii}{}/\Halpha{} line-ratios typical of AGN \citep{2001ApJ...556..121K}.
Unsurprisingly our model is unable the infer the metallicity profiles of galaxies with AGN. 
So we reiterate that when applying our method we must be careful to exclude such galaxies.

We conclude that, in general, our model is able to recover central metallicities and metallicity gradients from realistic galaxies.
However, while most galaxies lie close to the 1:1 line, a few of the galaxies with the steepest true metallicity gradients do not.
Several of these exhibit large differences between our two methods for defining the true metallicity gradient, clearly indicating that a metallicity gradient is poorly defined in these galaxies.
Nevertheless, there are a few galaxies for which our model significantly underestimates the metallicity gradient.
These, alongside NGC4980, could be considered as cases where our model fails catastrophically.

\subsection{Interpreting the observed metallicity gradient}

Our analysis has highlighted some intrinsic limitations when working with low-resolution data.
Namely the effect that clumpy emission will have on the inferred gradient, particularly if the clumps have uncharacteristically low/high metallicities.
This will become an important consideration if one is to compare the metallicity gradients of galaxies between the low and high redshift universes.

As mentioned in the introduction, there have been many reports of inverted (positive) metallicity gradients in high-redshift galaxies.
This is often interpreted as either evidence of possible accretion of metal poor gas to the centres of galaxies, or evidence for centrally concentrated winds which entrain metals in the outflow.
Therefore it is intriguing that a galaxy like NGC4980 that has a normal (negative) metallicity gradient can appear to have an inverted (positive) one when analysed using the methodology normally applied to distant galaxies.
It would be inappropriate for us to claim that clumpy emission explains any or all of the observed positive metallicity gradients.
However, we suggest that when interpreting these results, it is important to consider the implication that the positive gradients can be caused by low-metallicity strongly star-forming clumps, whose metallicity is not indicative of the overall metallicity profile.

In this section we have shown that our model performs satisfactorily well in both ideal and realistic scenarios.
Our model is able to recover the metallicity gradients of barely resolved galaxies, but we have identified that there are important considerations to be made with regards to the interpretation.
In the following section we will apply our method to real observations as a proof on concept.

\section{Application}\label{sec:application}

In the previous section we successfully tested our model against mock data.
We shall now demonstrate the model applied to real IFS observations of high-redshift galaxies.
This will allow us to assess how well the model can constrain the metallicity profile of distant galaxies.

\subsection{Data}

We will use MUSE observations of the Hubble Deep Field South (HDFS) which were taken during the last commissioning phase of MUSE (June--August 2014).
MUSE is an integral field spectrograph providing continuous spatial coverage over a $1\arcmin \times 1\arcmin$ FoV, across the wavelength range 4750\AA{} -- 9300\AA{}, with a spectral resolution of 2.3\AA{} FWHM.

The data and its reduction (version 1.0)\footnote{Public data products and catalogues are available at \url{http://muse-vlt.eu/science/}}are described at length by \citet{2015A&A...575A..75B}.
With the 54 exposures (27h) it is possible to obtain a $1\sigma$ emission-line surface-brightness limit of $1 \times 10^{-19}\,\mathrm{erg}\,\mathrm{s}^{-1}\,\mathrm{cm}^{-2}\,\mathrm{arcsec}^{-2}$.
Here we use a more recent reduction (version 1.24) that incorporates some minor improvements in the uniformity and sky subtraction of the data.
However, for the sources that we concern ourselves with here these modifications are not important.
The PSF in these observations is characterized by a Moffat profile with parameters as given in Table~\ref{tab:psf_params}.
The final data cube is sampled with equally sized voxels\footnote{volumetric pixels} ($0.2'' \times 0.2'' \times 1.25\angstrom$).

Our model requires a set of predetermined morphological parameters: the location of the centre of the galaxy, its inclination and the position angle of the major axis (PA).
The details of the measurement of these quantities are given in \citet{2016A&A...591A..49C}, but briefly they were determined by running GALFIT \citep{2002AJ....124..266P} on the F814W HST images \citep{1996AJ....112.1335W}, using a disc+bulge model.

We adopt the redshifts of the galaxies as those tabulated by \citet{2015A&A...575A..75B}.
We will also use the same object ID numbers.

\subsection{Analysis}

To separate the nebular emission from the underlying stellar component we do full-spectral fitting using the \textsc{platefit} code described in \citet{2004ApJ...613..898T} and \citet{2004MNRAS.351.1151B}. 
We process a spectrum as follows:
\begin{description}

\item[\textit{Redshift determination.}] Although we already know the redshift of each galaxy, the galaxy's own rotation will result in small velocity offsets from this value.
We determine the redshift of the spectrum using the \textsc{autoz} code described by \citet{2014MNRAS.441.2440B}, which determines redshifts using cross-correlations with template spectra.
If there is a strong correlation peak within $\pm500\,\mathrm{km}\,\mathrm{s}^{-1}$ of the galaxy's redshift, then we accept this peak as the redshift of the spectrum.
If no significant correlation peak is found within this range, we assume the spectrum's redshift to be the same as the galaxy as a whole.

\item[\textit{Stellar velocity dispersion.}] The stellar velocity dispersion is determined using \textsc{vdispfit}\footnote{\url{http://spectro.princeton.edu/idlspec2d_install.html}}.
This uses a set of eigenspectra, convolved for different velocity dispersions.
From this the best fit velocity dispersion is determined.
This value includes the instrumental velocity dispersion.
If the best fit velocity dispersion lies outside the range $\left[10 - 300\right]\,\mathrm{km}\,\mathrm{s}^{-1}$ we assume the fit has failed and adopt a default value of $80\,\mathrm{km}\,\mathrm{s}^{-1}$. 
Such failures are typical when the stellar continuum is faint or non-existent.

\item[\textit{Continuum fitting.}] For the spectral fitting we use the \textsc{platefit} spectral-fitting routine \citep{2004ApJ...613..898T, 2004MNRAS.351.1151B}.
\textsc{platefit}, which was developed for the SDSS, fits the stellar continuum and emission lines separately.
In this continuum fitting stage, regions around possible emission-lines are masked out.
The stellar continuum is fit with a collection of \citet{2003MNRAS.344.1000B} stellar population synthesis model templates.
The template fit is performed using the previously derived redshift and velocity dispersion.
If the continuum fitting fails, i.e. because the continuum has very low S/N, then we construct the continuum from a running-median filter with a 150\AA{} width.

\item[\textit{Emission-line fitting.}] The second \textsc{platefit} emission-line fitting stage is now performed on the residual spectrum (after continuum subtraction).
The emission lines are each modelled with a single Gaussian component.
Doublets such as \forbidden{O}{ii}{3726,3729} are fit with two Gaussian components.
All emission lines share a common velocity offset and a common velocity dispersion.
The velocity offset and velocity dispersion are not fixed, but are instead free parameters in the fit.
The amplitudes and associated errors are determined as part of a Levenberg-Marquardt least-squares minimization.
However, analysis of duplicate SDSS observations has shown that these formal errors typically underestimate the true uncertainties.
Corrections for this can, however, be derived from the duplicate observations \citep[e.g.][]{2013MNRAS.432.2112B}.
We use these corrections to rescale our formal uncertainties to more representative values.

\end{description}

For this paper we make it a requirement that all our emission-line flux measurements have $\textrm{S/N} \ge 5$.
Near the bright centres of galaxies individual spaxels will satisfy this criterion.
However, at larger radii we need to coadd spaxels to reach the required S/N.
To combat the effects of seeing we will need as much radial information as possible, and therefore it is necessary to bin (aggregate) spaxels together.
There is, however, no perfect binning algorithm.
We present our adopted procedure in Appendix~\ref{sec:binning_algorithm}.
The method bins the galaxy into annular sectors, and attempts to avoid binning spaxels at very different radii, although this last point is far from guaranteed.
This should help minimize addition radial resolution loss as a result of the binning.
It should be noted that these bins are not contiguous, i.e. non-adjacent spaxels will be combined.
In many cases the bins will be smaller than the PSF, and therefore the derived fluxes will not be statistically independent of one another.

\subsection{Results}\label{sec:application_results}

In this section we present the results of fitting our model to real data.
Using this we will discuss characteristics of the method, outline certain limitations, and discuss future improvements that could be made.

\begin{table}
\caption{Galaxy properties: disc scale-length, stellar mass and star-formation rate. These results were reported in \citet{2016A&A...591A..49C}, but we reproduce them here for convenience.}
\label{tab:contini_results}
\begin{tabular}
{c
d{1.3}@{\hspace{0.2em}}c@{\hspace{0.2em}}d{1.3}
d{1.2}@{\hspace{0.2em}}c@{\hspace{0.2em}}d{1.2}
d{2.2}@{\hspace{0.2em}}c@{\hspace{0.2em}}d{1.2}}
\hline
Galaxy &
\multicolumn{3}{c}{$r_d$} &
\multicolumn{3}{c}{$\log_{10}\left(\textrm{M}_\ast\right)$} &
%\multicolumn{3}{c}{\multirow{2}{*}{$\log_{10}\left(\displaystyle{\frac{\textrm{M}_\ast}{\textrm{M}_{\sun{}}}}\right)$}} &
\multicolumn{3}{c}{$\log_{10}\left(\textrm{SFR}\right)$} \\
%\multicolumn{3}{c}{\multirow{2}{*}{$\log_{10}\left(\displaystyle{\frac{\textrm{SFR}}{\textrm{M}_{\sun{}}\,\mathrm{yr}^{-1}}}\right)$}} \\
 &
\multicolumn{3}{c}{$[\textrm{arcsec}]$} & 
\multicolumn{3}{c}{$[\textrm{M}_{\sun{}}]$} & 
\multicolumn{3}{c}{$[\textrm{M}_{\sun{}}\,\mathrm{yr}^{-1}]$} \\
\hline
HDFS-0003 & 0.660 & $\pm$ & 0.007 & 9.66 & $\pm$ & 0.14 &  0.24 & $\pm$ & 0.37 \\
HDFS-0016 & 0.40  & $\pm$ & 0.01  & 8.74 & $\pm$ & 0.21 & -0.65 & $\pm$ & 0.55 \\
\hline
\end{tabular}
\end{table}

As examples we will show results for two galaxies, one of which is well resolved (HDFS-0003), and another barely resolved galaxy (HDFS-0016).
These galaxies were selected to represent these two extremes.
Of the two, HDFS-0003 is the larger, more massive, and more strongly star forming (see Table~\ref{tab:contini_results}).
Both galaxies have similar redshifts ($z=0.5637$ and $z=0.4647$, respectively), which means that the intrinsic physical resolution of both observations is approximately $4\,\textrm{kpc}$ FWHM.
In our analysis we use the same set of emission lines for both galaxies.

\begin{figure*}
\includegraphics[width=\linewidth]{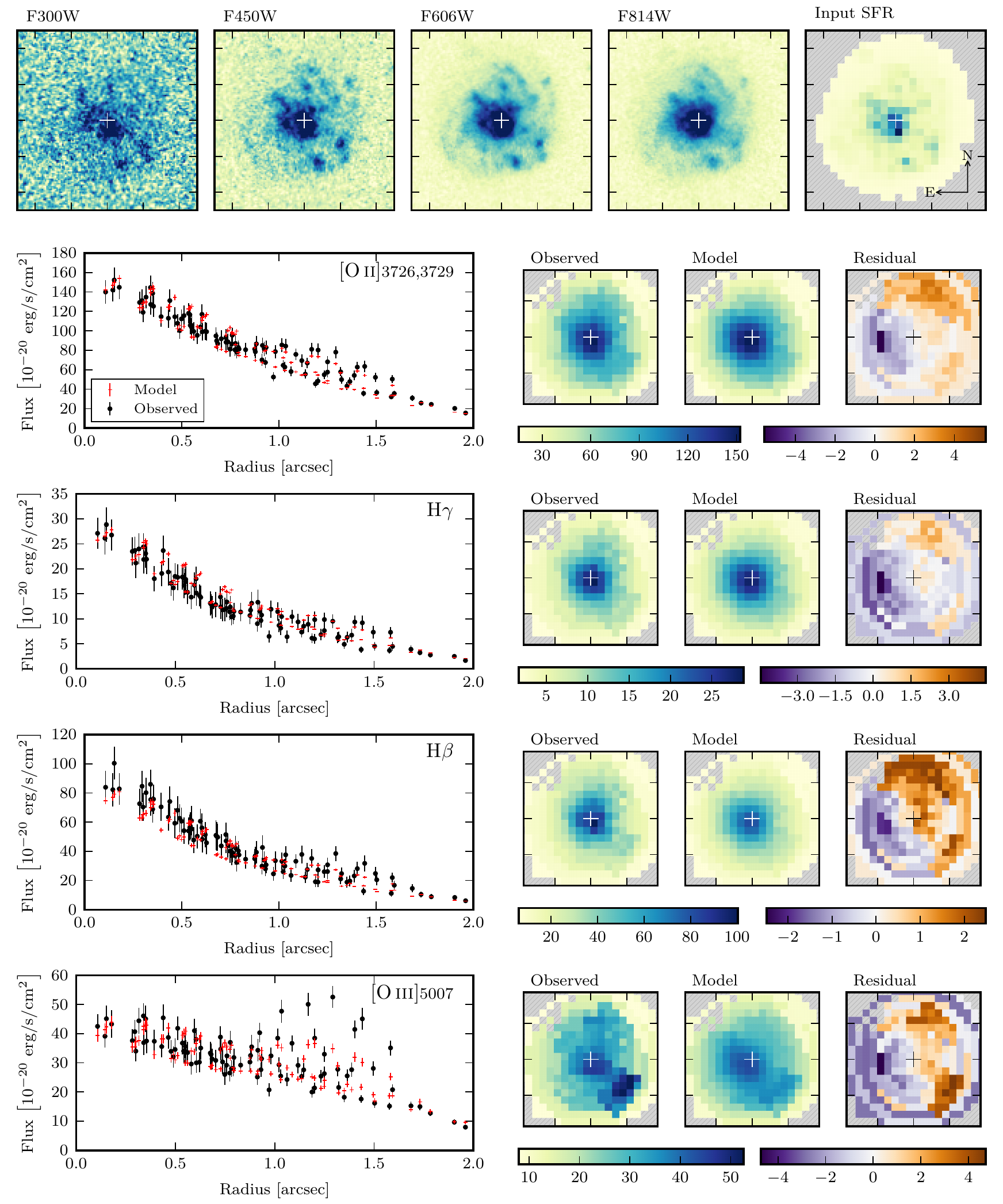}
\caption{Summary of model fitting for visual quality assessment of galaxy HDFS-0003.
(Top) We plot five images: four HST broadband images, and the derived SFR map which is used as an input to the model.
(Left) We show the radial flux profiles for all four emission-lines (\forbidden{O}{ii}{}, \Hgamma{}, \Hbeta{} and \forbidden{O}{iii}{}).
Black data points indicate observed fluxes and their $\pm1\sigma$ errors.
The red crosses show the median model solution, the size of the vertical bar indicates a $\pm2\sigma$ range in fluxes.
(Right) Three images respectively show 2D binned images of the observed fluxes, model fluxes, and scaled residuals ($\sfrac{\textrm{\larger{}(Observed} - \textrm{\larger{}Model)}}{\textrm{\larger{}Error}}$) for each emission line.} 
\label{fig:model_fits_0003}
\end{figure*}

\begin{figure*}
\includegraphics[width=\linewidth]{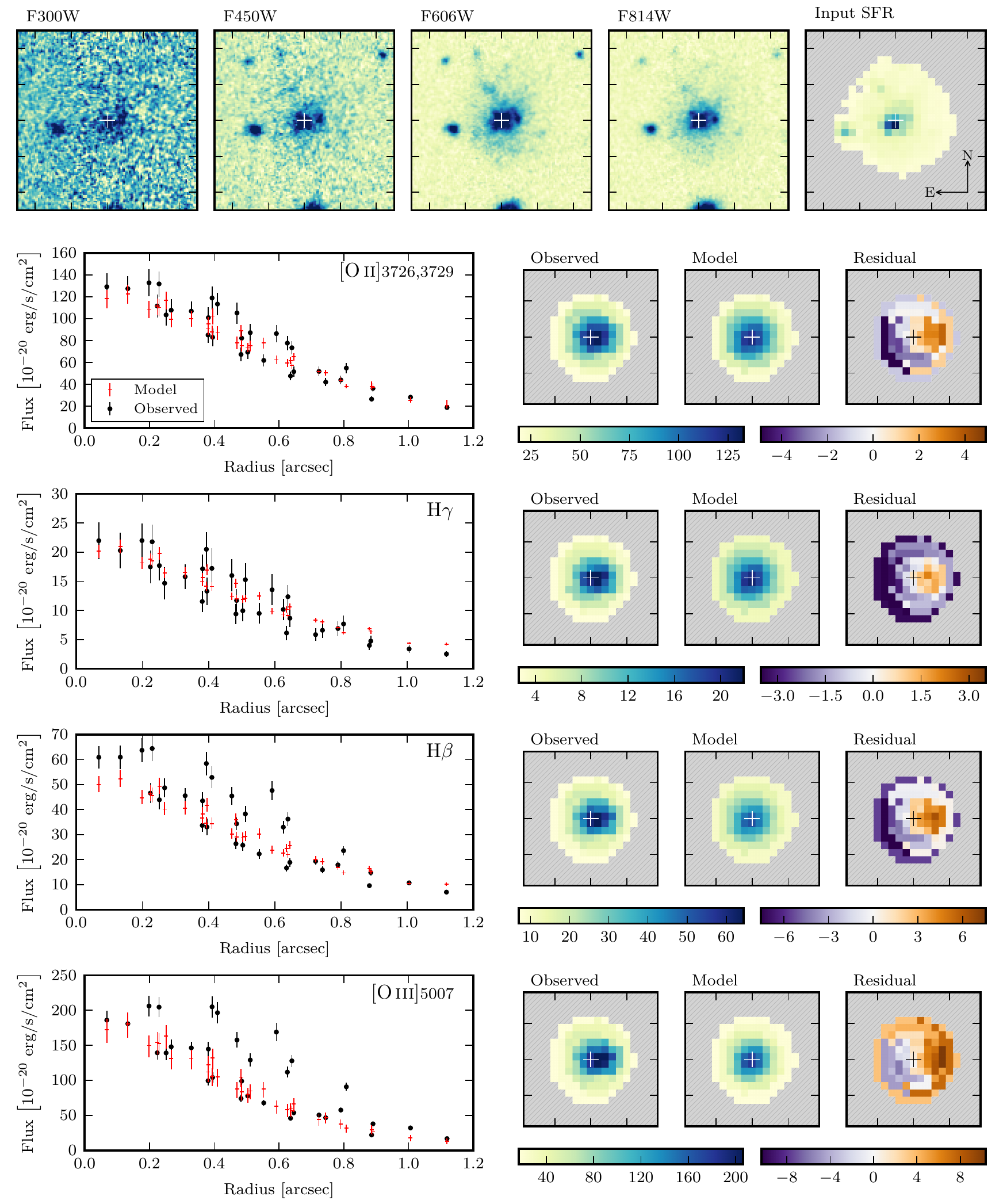}
\caption{Summary of model fitting for visual quality assessment of galaxy HDFS-0016.
See Fig.~\ref{fig:model_fits_0003} for details.}
\label{fig:model_fits_0016}
\end{figure*}

\begin{figure*}
\includegraphics[width=\linewidth]{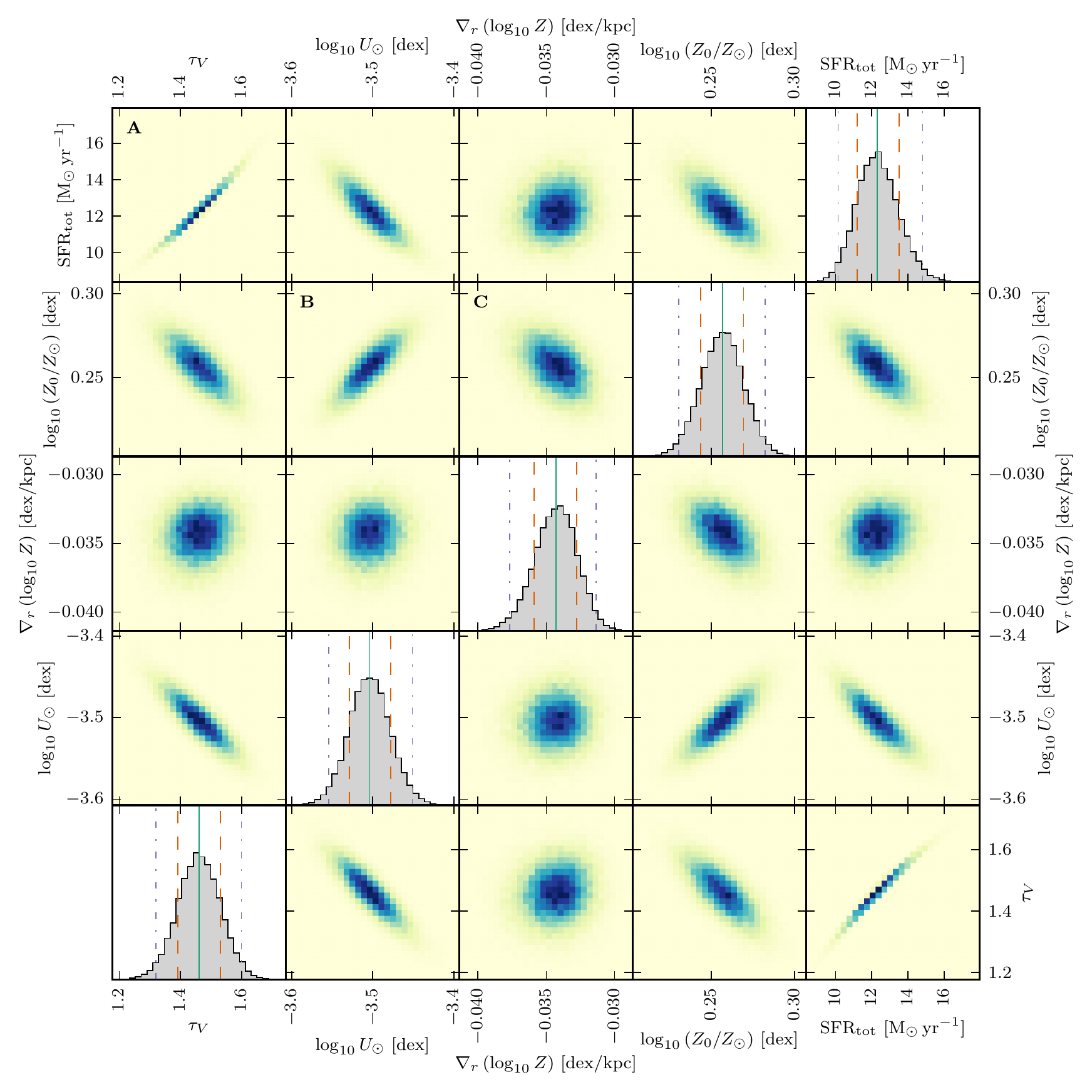}
\caption{MCMC fitting results shown for galaxy HDFS-0003.
We show both 1D and 2D marginalized histograms for all 5 parameters: the total star-formation rate, $\textrm{SFR}_\mathrm{tot}$, central metallicity, $\log_{10}Z_0$, metallicity gradient, $\nabla_r\left(\log_{10}Z\right)$, ionization parameter at solar metallicity, $\log_{10}U_{\sun}$, and V-band optical depth, $\tau_V$.
In each 1D histogram the vertical lines indicate the median (solid), $\pm1\sigma$ quantiles (dashed) and $\pm2\sigma$ quantiles (dash-dotted).
All axes span a $[-4\sigma,4\sigma]$ interval in their respective parameters.
Letters label particular panels that we refer to in the text.}
\label{fig:triangle_0003}
\end{figure*}

\begin{figure*}
\includegraphics[width=\linewidth]{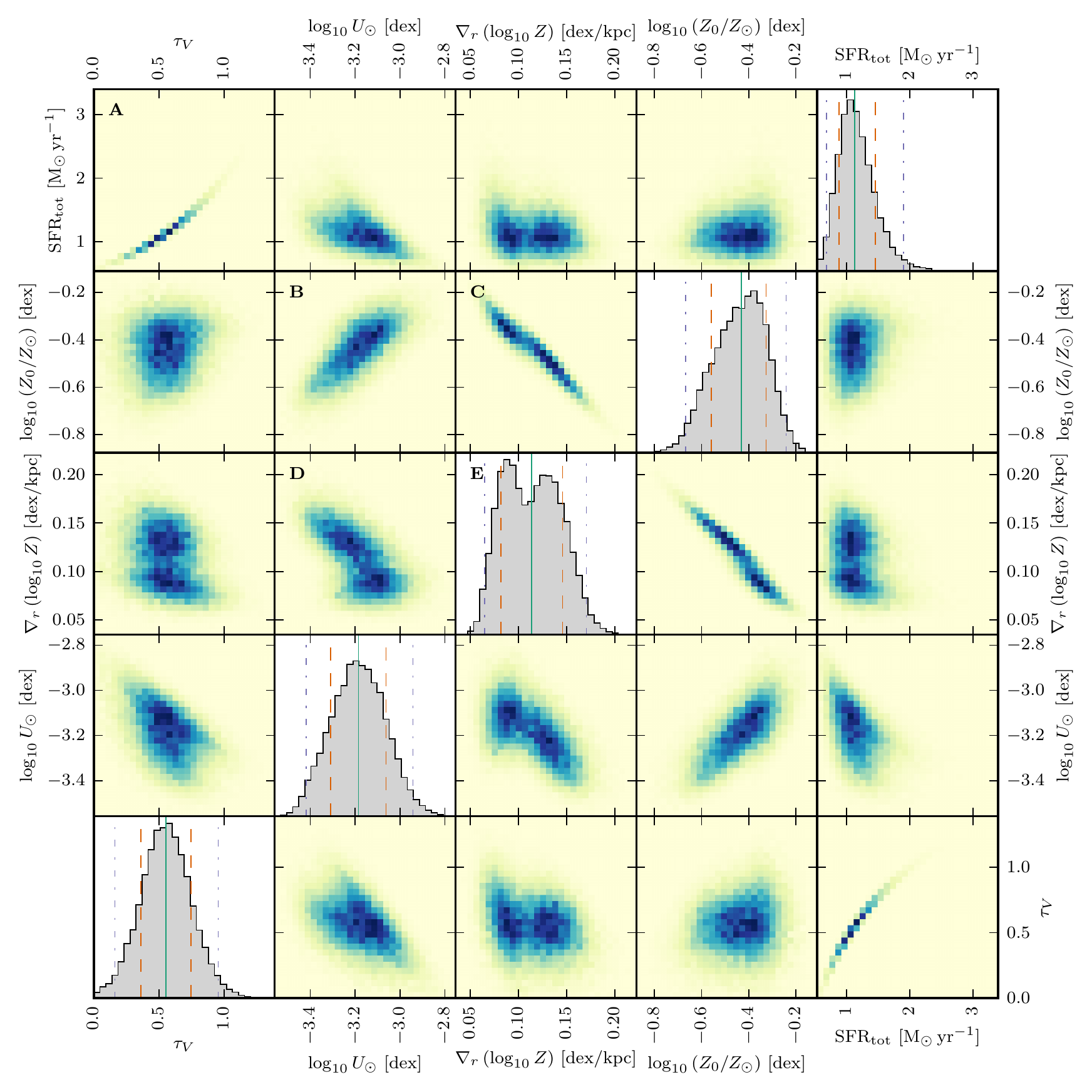}
\caption{MCMC fitting results shown for galaxy HDFS-0016.
See Fig.~\ref{fig:triangle_0003} for details.}
\label{fig:triangle_0016}
\end{figure*}

In Figs.~\ref{fig:model_fits_0003}~\&~\ref{fig:model_fits_0016} we present a comparison between the observed emission-line fluxes and the model fit for the two galaxies.
The model reproduces the observed emission-line fluxes in both.
However, while the model is able to capture the overall radial flux profile, it does not (by construction) have the flexibility to match the observed azimuthal metallicity variations.
This is especially evident in HDFS-0016 where the emission line fluxes are not single-valued at all radii.
In this galaxy it appears that the radial run of emission line fluxes could by described by two branches, with the brightest branch originating from a star-forming clump offset to the West of the galaxy centre.

We discussed in the previous section that star-forming clumps can conceptually be divided into two categories: either clumps that are bright, but have the same metallicity as other gas at the same radius, and clumps which have uncharacteristically low/high metallicities.
In the case of the former the line ratios (but not line fluxes) would be single valued as function of radius.\footnote{This is not entirely true since seeing has a wavelength dependence.}
However, HDFS-0016 falls into the latter category as it is clear to the eye that the upper branch of fluxes has a consistently higher \mbox{\forbidden{O}{iii}{}/\forbidden{O}{ii}{}} ratio.
For a range of radii in HDFS-0016 there is no single characteristic line-ratio.

The existence of multiple branches in the flux profiles can cause problems for the model fitting even if the line-ratios are unaltered.
One can envisage a scenario where, for example, the model might have fit the upper branch in \forbidden{O}{iii}{}, but the lower branch in \forbidden{O}{ii}{}.
Obviously this would result in deriving an entirely incorrect best fit model.
Indeed Fig.~\ref{fig:model_fits_0016} shows slight hints of this problem.
Notably the model fits the lower flux branch in all emission lines, except for \Hgamma{} where the model fits in between the lower and upper branches.
Albeit relatively minor in this case, it is crucial to be aware of this possible problem and assess its severity.

\subsubsection{Validity of constant dust approximation}\label{sec:application_results_dust}

For our model we assume there is a constant attenuation due to dust across the whole galaxy.
Studying the \Hbeta{} and \Hgamma{} profiles in both Figs.~\ref{fig:model_fits_0003}~\&~\ref{fig:model_fits_0016} one observes that the model slightly underpredicts the \Hbeta{} flux in the centre of the galaxies.
This would imply that there is perhaps a mild dust gradient across the galaxy, with galaxy centres being slightly more dusty than their outskirts.

Using high spatial-resolution grism spectroscopy \citet{2016ApJ...817L...9N} identified radial dust variations in z=1.4 galaxies.
They found that the most massive galaxies presented the strongest variations, but less massive $10^{9.2}\,\textrm{M}_{\sun{}}$ galaxies exhibited almost no variation and little dust attenuation overall.

We reperform our analysis of HDFS-0003 using a dust model with the same radial dependence as \citet{2016ApJ...817L...9N} propose for a $10^{9.66}\,\textrm{M}_{\sun{}}$ galaxy.
The normalization of this model is allowed as a free parameter.
We find that dust model produces a significantly worse fit to the data than the constant dust model.
Admittedly, since the \citet{2016ApJ...817L...9N} dust models are based on z=1.4 galaxies they may not be appropriate for our galaxies.

Using the new dust model changes many of the derived best fit values.
For example the inferred central metallicity is increased by $\sim0.14\,\textrm{dex}$, however, the metallicity gradient is bizarrely unaffected and changes by $<0.001\,\textrm{dex}/\textrm{kpc}$.

Choosing a appropriate dust model is clearly important for deriving the metallicity of galaxies.
But, on the whole the data appears largely consistent with our assumption of a constant optical depth for the whole galaxy.

\subsubsection{Parameter constraints}

So far we have only discussed the quality of the model fits.
We will now discuss how well the model can constrain the metallicity profile of these galaxies.
In Figs.~\ref{fig:triangle_0003}~\&~\ref{fig:triangle_0016} we show 1D and 2D histograms of the derived model parameters for both HDFS-0003 and HDFS-0016.
We note that most of the derived parameters are relatively well constrained.
For example in HDFS-0016 the errors on central metallicity and metallicity gradient are $\pm0.1\ \textrm{dex}$ and $\pm0.03\ \textrm{dex}/\textrm{kpc}$ respectively.
These errors are more than sufficient to establish HDFS-0016 as possessing a significantly sub-solar central metallicity and a positive metallicity gradient.
The constraints on HDFS-0003 are tighter.
Naturally the quality of the constraints will vary with the S/N of the data.
It is therefore perhaps more interesting to discuss the correlations between the modelled parameters.

It is clear from Panel~A that the model produces a very tight correlation between the total star-formation rate and the V-band optical depth.
Dustier model solutions are fainter, so intrinsically higher SFRs are required to compensate.

The model also shows a strong anti-correlation between the metallicity gradient and central metallicity of a galaxy (Panel~C).
This degeneracy is of course not surprising given that data directly constrains the metallicity profile, not the metallicity gradient, which is dependent on the central metallicity.
However, the situation may actually be more complicated than this.
For example in HDFS-0016, as depicted by Panels~B \& D, the ionization parameter at solar metallicity, $\log_{10}U_{\sun{}}$, is (anti-)correlated with both the metallicity gradient and central metallicity.
HDFS-0003 does not show this dependence between $\log_{10}U_{\sun{}}$ and $\nabla_r\left(\log_{10}Z\right)$.
However, HDFS-0003 does show an interdependency between $\textrm{SFR}_\mathrm{tot}$, $\log_{10}Z_0$, $\log_{10}U_{\sun{}}$ and $\tau_V$.

Because we have assumed an intrinsic correlation between metallicity and ionization parameter, it is somewhat difficult to unravel these dependencies.
In essence metallicity gradients and ionization-parameter gradients are one and the same.
It is this which allows us to mitigate against the $\textrm{R}_{23}$ degeneracy (the degeneracy between metallicity and ionization parameter that arises from the limited set of emission lines used here).
However, as a consequence the ionization parameter at solar metallicity, central metallicity and metallicity gradient are now inadvertently coupled.

Interestingly, the metallicity gradient in HDFS-0016 is slightly bimodal (see Panel~E).
An effect which may in part be explained by the dual-valued nature of the $\textrm{R}_{23}$ degeneracy, although this is hard to verify.
Currently there is insufficient evidence to place an informative prior on $\log_{10}U_{\sun{}}$.
If in the future this were possible one could in theory achieve a more precise measurement for central metallicity and the metallicity gradient.

It is important to note that in our model testing we have only verified the central metallicity and the metallicity gradient parameters.
We have not applied the same testing scrutiny to the other parameters, so their values should not be considered validated and used only with great care.

\subsection{Discussion}

As we have seen, we can use the model to constrain the true metallicity gradient in galaxies.
To emphasise the necessity for correcting for the effects of seeing, we have also derived the metallicity profiles of these galaxies without making any corrections for seeing.

\begin{table}
\caption{Comparison of derived metallicity profile parameters from two methods.
One method is a simple linear fit to the metallicity derived in a series of annular bins.
The other is the full model fitting that accounts for seeing effects.}
\label{tab:simple_profile}
\begin{tabular}{cccc}
\hline
\multirow{2}{*}{Galaxy} & \multirow{2}{*}{Parameter} & Simple & Full \\
& & Annular & Modelling \\
\hline
\multirow{4}{*}{HDFS-0003} & $\log_{10}\left(Z_0/Z_{\sun{}}\right)$ &
\multirow{2}{*}{$+0.31 \pm 0.01$} & \multirow{2}{*}{$+0.26^{+0.01}_{-0.01}$} \\
 & ${\scriptstyle[\textrm{dex}]}$ & & \\
\cline{3-4}
 & $\nabla_r\log_{10}\left(Z\right)$ &
\multirow{2}{*}{$-0.026 \pm 0.002$} & \multirow{2}{*}{$-0.034^{+0.001}_{-0.002}$} \\
 & ${\scriptstyle[\textrm{dex}/\textrm{kpc}]}$ & & \\
\hline
\multirow{4}{*}{HDFS-0016} & $\log_{10}\left(Z_0/Z_{\sun{}}\right)$ &
\multirow{2}{*}{$-0.28 \pm 0.02$} & \multirow{2}{*}{$-0.43^{+0.10}_{-0.13}$} \\
 & ${\scriptstyle[\textrm{dex}]}$ & & \\
\cline{3-4}
 & $\nabla_r\log_{10}\left(Z\right)$ &
\multirow{2}{*}{$+0.016 \pm 0.004$} & \multirow{2}{*}{$+0.11^{+0.03}_{-0.03}$} \\
 & ${\scriptstyle[\textrm{dex}/\textrm{kpc}]}$ & & \\
\hline
\end{tabular}
\end{table}

We extract emission-line fluxes in a series of elliptical annular apertures (semi-major width $0.35''$) with axis-ratios to match the galaxy.
In each annulus we derive the metallicity following \citet[][herein M08]{2008A&A...488..463M}\defcitealias{2008A&A...488..463M}{M08}, except that our method differs slightly as we use the \citet{2000ApJ...539..718C} dust absorption model.
We use the same set of emission lines as for the full modelling, but also include the \forbidden{O}{iii}{4959} required for the $\textrm{R}_{23}$ index.

In Table~\ref{tab:simple_profile} we summarize the derived central metallicities and gradients, and compare them to those derived from the full modelling.
As a cautionary note it can be dangerous to compare metallicities derived from different methods and calibrations (\citet{2008ApJ...681.1183K} provide a good discussion of this).
Nevertheless it is still interesting to compare the results, as they should be broadly consistent.

HDFS-0003 is a well resolved galaxy, therefore the effects of seeing will be limited.
Indeed both methods produce shallow, negative metallicity gradients.
Although the annular method is slightly shallower, this is not likely to be seeing effect and is more probably due to differences between the methods for deriving metallicity and/or the fact that the annular method derives the dust in each annulus, allowing for possible radial dust variations.

In stark contrast, HDFS-0016 will be much more affected by seeing effects.
The predominant effects of seeing will be to flatten the metallicity gradient.
And this is exactly what is observed, the na\"{i}ve annular method yields a significantly flatter (but still positive) metallicity gradient.
This method also estimates a $\sim 0.15\,\textrm{dex}$ higher central metallicity.
While this could entirely be due to difference between the methods for deriving metallicity, there are other important factors to consider.
If the galaxy truly has a steep positive metallicity gradient, then a significant fraction of the flux from the outer, higher metallicity material could be scattered into the central bin.
Thus the uncorrected central metallicity may be much closer to the average metallicity of the galaxy (although given the non-linear nature of the connection between metallicity and emission-line flux this may not necessarily be the case in all galaxies).

As a final cautionary note, throughout this section we have made use of high-resolution SFR maps to provide a more realistic model for these galaxies.
Whilst employing SFR maps may be theoretically optimal, in practice good SFR maps are challenging to obtain.
The SFR maps contain systematic and random errors.
For example in Fig.~\ref{fig:model_fits_0016} we observe a star-forming clump to the West of the galaxy centre which is seen in the HST images. 
This clump is however not apparent in the derived SFR maps.
Additionally the SFR maps can be contaminated by other galaxies in the (fore/back)ground.
Both the systematic and random uncertainties, which are not factored into the modelling, may limit or even negate their effectiveness.

\section{Conclusions}\label{sec:conclusions}

It is important to correct for the effects of seeing when determining metallicity gradients in galaxies.
Here we have outlined an approach that allows us to directly model the emission-line fluxes.
By fitting this model to the data we can infer the true metallicity profile of a galaxy in the absence of seeing.
Unlike other existing approaches, our method is general can be applied to many IFS studies of distant galaxies.

We use theoretical photoionization models to predict the emission-line ratios as a function of metallicity and ionization parameter.
As such the model can be applied to a flexible set of observed emission-lines, enabling a self-consistent analysis across a range of redshifts and a variety of instrument wavelength coverages.
To alleviate degeneracies we enforce a correlation between metallicity and the ionization parameter.
We, however, do permit global ionization-parameter variations, accommodating for both possible redshift and environmental evolution of the ionization parameter.

We have performed an extensive set of tests to validate the method and understand its limitations.
In summary:
\begin{enumerate}

\item By creating noisy model realizations for a variety of S/N, inclination, and seeing conditions, we have established that the model is able to recover the metallicity profile even in adverse conditions.
In addition the method produces appropriate error estimates.

\item We have downgraded observations of nearby galaxies to test our method against realistic mock data.
With limited resolution the metallicity profile will inevitably be weighted towards the metallicity of the brightest clumps.

\item This effect is not wholly reversible, even if the underlying SFR distribution is known a priori.
Providing a good map of the underlying SFR distribution is challenging, and proves to be the greatest limitation for our model.

\item The ability for bright star-forming clumps to skew the measured metallicity gradient should be taken into account when interpreting metallicity gradient studies.

\end{enumerate}

In future work we will apply this method to allow us derive the metallicity profiles of galaxies observed with MUSE.

\section*{Acknowledgements}

JB acknowledges support from FCT grant IF/01654/2014/CP1215/CT0003. 
TC acknowledges support from the ANR FOGHAR (ANR-13-BS05-0010-02), the OCEVU Labex (ANR-11-LABX-0060) and the A*MIDEX project (ANR-11-IDEX-0001-02) funded by the ``Investissements d'avenir'' French government program managed by the ANR.
BE acknowledges financial support from ''Programme National de Cosmologie and Galaxie'' (PNCG) of CNRS/INSU, France.
TPKM acknowledges financial support from the Spanish Ministry of Economy and Competitiveness (MINECO) under grant number AYA2013-41243-P.

Funding for the SDSS and SDSS-II has been provided by the Alfred P. Sloan Foundation, the Participating Institutions, the National Science Foundation, the U.S. Department of Energy, the National Aeronautics and Space Administration, the Japanese Monbukagakusho, the Max Planck Society, and the Higher Education Funding Council for England. The SDSS Web Site is \url{http://www.sdss.org}.

The SDSS is managed by the Astrophysical Research Consortium for the Participating Institutions. The Participating Institutions are the American Museum of Natural History, Astrophysical Institute Potsdam, University of Basel, University of Cambridge, Case Western Reserve University, University of Chicago, Drexel University, Fermilab, the Institute for Advanced Study, the Japan Participation Group, Johns Hopkins University, the Joint Institute for Nuclear Astrophysics, the Kavli Institute for Particle Astrophysics and Cosmology, the Korean Scientist Group, the Chinese Academy of Sciences (LAMOST), Los Alamos National Laboratory, the Max-Planck-Institute for Astronomy (MPIA), the Max-Planck-Institute for Astrophysics (MPA), New Mexico State University, Ohio State University, University of Pittsburgh, University of Portsmouth, Princeton University, the United States Naval Observatory, and the University of Washington.

We acknowledge the usage of the HyperLeda database (\url{http://leda.univ-lyon1.fr}).

This study uses data provided by the Calar Alto Legacy Integral Field Area (CALIFA) survey (\url{http://califa.caha.es}).
Based on observations collected at the Centro Astron\'{o}mico Hispano Alem\'{a}n (CAHA) at Calar Alto, operated jointly by the Max-Planck-Institut f\:{u}r Astronomie and the Instituto de Astrof\'{i}sica de Andaluc\'{i}a (CSIC).

Additionally we wish to acknowledge both the \textsc{Python} programming language and the Interactive Data Language (\textsc{idl}) that were both used extensively throughout this work.

%%%%%%%%%%%%%%%%%%%%%%%%%%%%%%%%%%%%%%%%%%%%%%%%%%



\bibliographystyle{mnras}
\bibliography{references.bib}

\begin{thebibliography}{}
\makeatletter
\relax
\def\mn@urlcharsother{\let\do\@makeother \do\$\do\&\do\#\do\^\do\_\do\%\do\~}
\def\mn@doi{\begingroup\mn@urlcharsother \@ifnextchar [ {\mn@doi@}
  {\mn@doi@[]}}
\def\mn@doi@[#1]#2{\def\@tempa{#1}\ifx\@tempa\@empty \href
  {http://dx.doi.org/#2} {doi:#2}\else \href {http://dx.doi.org/#2} {#1}\fi
  \endgroup}
\def\mn@eprint#1#2{\mn@eprint@#1:#2::\@nil}
\def\mn@eprint@arXiv#1{\href {http://arxiv.org/abs/#1} {{\tt arXiv:#1}}}
\def\mn@eprint@dblp#1{\href {http://dblp.uni-trier.de/rec/bibtex/#1.xml}
  {dblp:#1}}
\def\mn@eprint@#1:#2:#3:#4\@nil{\def\@tempa {#1}\def\@tempb {#2}\def\@tempc
  {#3}\ifx \@tempc \@empty \let \@tempc \@tempb \let \@tempb \@tempa \fi \ifx
  \@tempb \@empty \def\@tempb {arXiv}\fi \@ifundefined
  {mn@eprint@\@tempb}{\@tempb:\@tempc}{\expandafter \expandafter \csname
  mn@eprint@\@tempb\endcsname \expandafter{\@tempc}}}

\bibitem[\protect\citeauthoryear{{Abazajian} et~al.,}{{Abazajian}
  et~al.}{2009}]{2009ApJS..182..543A}
{Abazajian} K.~N.,  et~al., 2009, \mn@doi [\apjs]
  {10.1088/0067-0049/182/2/543}, \href
  {http://adsabs.harvard.edu/abs/2009ApJS..182..543A} {182, 543}

\bibitem[\protect\citeauthoryear{{Aller} \& {Liller}}{{Aller} \&
  {Liller}}{1959}]{1959ApJ...130...45A}
{Aller} L.~H.,  {Liller} W.,  1959, \mn@doi [\apj] {10.1086/146695}, \href
  {http://adsabs.harvard.edu/abs/1959ApJ...130...45A} {130, 45}

\bibitem[\protect\citeauthoryear{{Bacon} et~al.,}{{Bacon}
  et~al.}{2010}]{2010SPIE.7735E..08B}
{Bacon} R.,  et~al., 2010, in Ground-based and Airborne Instrumentation for
  Astronomy III. p. 773508, \mn@doi{10.1117/12.856027}

\bibitem[\protect\citeauthoryear{{Bacon} et~al.,}{{Bacon}
  et~al.}{2015}]{2015A&A...575A..75B}
{Bacon} R.,  et~al., 2015, \mn@doi [\aap] {10.1051/0004-6361/201425419}, \href
  {http://adsabs.harvard.edu/abs/2015A%26A...575A..75B} {575, A75}

\bibitem[\protect\citeauthoryear{{Baldry} et~al.,}{{Baldry}
  et~al.}{2014}]{2014MNRAS.441.2440B}
{Baldry} I.~K.,  et~al., 2014, \mn@doi [\mnras] {10.1093/mnras/stu727}, \href
  {http://adsabs.harvard.edu/abs/2014MNRAS.441.2440B} {441, 2440}

\bibitem[\protect\citeauthoryear{{Bigiel}, {Leroy}, {Walter}, {Brinks}, {de
  Blok}, {Madore}  \& {Thornley}}{{Bigiel} et~al.}{2008}]{2008AJ....136.2846B}
{Bigiel} F.,  {Leroy} A.,  {Walter} F.,  {Brinks} E.,  {de Blok} W.~J.~G.,
  {Madore} B.,   {Thornley} M.~D.,  2008, \mn@doi [\aj]
  {10.1088/0004-6256/136/6/2846}, \href
  {http://adsabs.harvard.edu/abs/2008AJ....136.2846B} {136, 2846}

\bibitem[\protect\citeauthoryear{{Blanc}, {Kewley}, {Vogt}  \&
  {Dopita}}{{Blanc} et~al.}{2015}]{2015ApJ...798...99B}
{Blanc} G.~A.,  {Kewley} L.,  {Vogt} F.~P.~A.,   {Dopita} M.~A.,  2015, \mn@doi
  [\apj] {10.1088/0004-637X/798/2/99}, \href
  {http://adsabs.harvard.edu/abs/2015ApJ...798...99B} {798, 99}

\bibitem[\protect\citeauthoryear{{Bouch{\'e}} et~al.,}{{Bouch{\'e}}
  et~al.}{2010}]{2010ApJ...718.1001B}
{Bouch{\'e}} N.,  et~al., 2010, \mn@doi [\apj] {10.1088/0004-637X/718/2/1001},
  \href {http://adsabs.harvard.edu/abs/2010ApJ...718.1001B} {718, 1001}

\bibitem[\protect\citeauthoryear{{Brinchmann}, {Charlot}, {White}, {Tremonti},
  {Kauffmann}, {Heckman}  \& {Brinkmann}}{{Brinchmann}
  et~al.}{2004}]{2004MNRAS.351.1151B}
{Brinchmann} J.,  {Charlot} S.,  {White} S.~D.~M.,  {Tremonti} C.,  {Kauffmann}
  G.,  {Heckman} T.,   {Brinkmann} J.,  2004, \mn@doi [\mnras]
  {10.1111/j.1365-2966.2004.07881.x}, \href
  {http://adsabs.harvard.edu/abs/2004MNRAS.351.1151B} {351, 1151}

\bibitem[\protect\citeauthoryear{{Brinchmann}, {Charlot}, {Kauffmann},
  {Heckman}, {White}  \& {Tremonti}}{{Brinchmann}
  et~al.}{2013}]{2013MNRAS.432.2112B}
{Brinchmann} J.,  {Charlot} S.,  {Kauffmann} G.,  {Heckman} T.,  {White}
  S.~D.~M.,   {Tremonti} C.,  2013, \mn@doi [\mnras] {10.1093/mnras/stt551},
  \href {http://adsabs.harvard.edu/abs/2013MNRAS.432.2112B} {432, 2112}

\bibitem[\protect\citeauthoryear{{Bruzual} \& {Charlot}}{{Bruzual} \&
  {Charlot}}{2003}]{2003MNRAS.344.1000B}
{Bruzual} G.,  {Charlot} S.,  2003, \mn@doi [\mnras]
  {10.1046/j.1365-8711.2003.06897.x}, \href
  {http://adsabs.harvard.edu/abs/2003MNRAS.344.1000B} {344, 1000}

\bibitem[\protect\citeauthoryear{{Buchner} et~al.,}{{Buchner}
  et~al.}{2014}]{2014A&A...564A.125B}
{Buchner} J.,  et~al., 2014, \mn@doi [\aap] {10.1051/0004-6361/201322971},
  \href {http://adsabs.harvard.edu/abs/2014A%26A...564A.125B} {564, A125}

\bibitem[\protect\citeauthoryear{{Cappellari} \& {Copin}}{{Cappellari} \&
  {Copin}}{2003}]{2003MNRAS.342..345C}
{Cappellari} M.,  {Copin} Y.,  2003, \mn@doi [\mnras]
  {10.1046/j.1365-8711.2003.06541.x}, \href
  {http://adsabs.harvard.edu/abs/2003MNRAS.342..345C} {342, 345}

\bibitem[\protect\citeauthoryear{{Charlot} \& {Fall}}{{Charlot} \&
  {Fall}}{2000}]{2000ApJ...539..718C}
{Charlot} S.,  {Fall} S.~M.,  2000, \mn@doi [\apj] {10.1086/309250}, \href
  {http://adsabs.harvard.edu/abs/2000ApJ...539..718C} {539, 718}

\bibitem[\protect\citeauthoryear{{Contini} et~al.,}{{Contini}
  et~al.}{2016}]{2016A&A...591A..49C}
{Contini} T.,  et~al., 2016, \mn@doi [\aap] {10.1051/0004-6361/201527866},
  \href {http://adsabs.harvard.edu/abs/2016A%26A...591A..49C} {591, A49}

\bibitem[\protect\citeauthoryear{{Cresci}, {Mannucci}, {Maiolino}, {Marconi},
  {Gnerucci}  \& {Magrini}}{{Cresci} et~al.}{2010}]{2010Natur.467..811C}
{Cresci} G.,  {Mannucci} F.,  {Maiolino} R.,  {Marconi} A.,  {Gnerucci} A.,
  {Magrini} L.,  2010, \mn@doi [\nat] {10.1038/nature09451}, \href
  {http://adsabs.harvard.edu/abs/2010Natur.467..811C} {467, 811}

\bibitem[\protect\citeauthoryear{{Dav{\'e}}, {Finlator}  \&
  {Oppenheimer}}{{Dav{\'e}} et~al.}{2012}]{2012MNRAS.421...98D}
{Dav{\'e}} R.,  {Finlator} K.,   {Oppenheimer} B.~D.,  2012, \mn@doi [\mnras]
  {10.1111/j.1365-2966.2011.20148.x}, \href
  {http://adsabs.harvard.edu/abs/2012MNRAS.421...98D} {421, 98}

\bibitem[\protect\citeauthoryear{{Diehl} \& {Statler}}{{Diehl} \&
  {Statler}}{2006}]{2006MNRAS.368..497D}
{Diehl} S.,  {Statler} T.~S.,  2006, \mn@doi [\mnras]
  {10.1111/j.1365-2966.2006.10125.x}, \href
  {http://adsabs.harvard.edu/abs/2006MNRAS.368..497D} {368, 497}

\bibitem[\protect\citeauthoryear{{Dopita} \& {Evans}}{{Dopita} \&
  {Evans}}{1986}]{1986ApJ...307..431D}
{Dopita} M.~A.,  {Evans} I.~N.,  1986, \mn@doi [\apj] {10.1086/164432}, \href
  {http://adsabs.harvard.edu/abs/1986ApJ...307..431D} {307, 431}

\bibitem[\protect\citeauthoryear{{Dopita}, {Kewley}, {Heisler}  \&
  {Sutherland}}{{Dopita} et~al.}{2000}]{2000ApJ...542..224D}
{Dopita} M.~A.,  {Kewley} L.~J.,  {Heisler} C.~A.,   {Sutherland} R.~S.,  2000,
  \mn@doi [\apj] {10.1086/309538}, \href
  {http://adsabs.harvard.edu/abs/2000ApJ...542..224D} {542, 224}

\bibitem[\protect\citeauthoryear{{Dopita} et~al.,}{{Dopita}
  et~al.}{2006}]{2006ApJ...647..244D}
{Dopita} M.~A.,  et~al., 2006, \mn@doi [\apj] {10.1086/505418}, \href
  {http://adsabs.harvard.edu/abs/2006ApJ...647..244D} {647, 244}

\bibitem[\protect\citeauthoryear{{Dopita}, {Sutherland}, {Nicholls}, {Kewley}
  \& {Vogt}}{{Dopita} et~al.}{2013}]{2013ApJS..208...10D}
{Dopita} M.~A.,  {Sutherland} R.~S.,  {Nicholls} D.~C.,  {Kewley} L.~J.,
  {Vogt} F.~P.~A.,  2013, \mn@doi [\apjs] {10.1088/0067-0049/208/1/10}, \href
  {http://adsabs.harvard.edu/abs/2013ApJS..208...10D} {208, 10}

\bibitem[\protect\citeauthoryear{{Feroz} \& {Hobson}}{{Feroz} \&
  {Hobson}}{2008}]{2008MNRAS.384..449F}
{Feroz} F.,  {Hobson} M.~P.,  2008, \mn@doi [\mnras]
  {10.1111/j.1365-2966.2007.12353.x}, \href
  {http://adsabs.harvard.edu/abs/2008MNRAS.384..449F} {384, 449}

\bibitem[\protect\citeauthoryear{{Feroz}, {Hobson}  \& {Bridges}}{{Feroz}
  et~al.}{2009}]{2009MNRAS.398.1601F}
{Feroz} F.,  {Hobson} M.~P.,   {Bridges} M.,  2009, \mn@doi [\mnras]
  {10.1111/j.1365-2966.2009.14548.x}, \href
  {http://adsabs.harvard.edu/abs/2009MNRAS.398.1601F} {398, 1601}

\bibitem[\protect\citeauthoryear{{Feroz}, {Hobson}, {Cameron}  \&
  {Pettitt}}{{Feroz} et~al.}{2013}]{2013arXiv1306.2144F}
{Feroz} F.,  {Hobson} M.~P.,  {Cameron} E.,   {Pettitt} A.~N.,  2013, preprint,
  \href {http://adsabs.harvard.edu/abs/2013arXiv1306.2144F} {} (\mn@eprint
  {arXiv} {1306.2144})

\bibitem[\protect\citeauthoryear{{Gallazzi}, {Charlot}, {Brinchmann}, {White}
  \& {Tremonti}}{{Gallazzi} et~al.}{2005}]{2005MNRAS.362...41G}
{Gallazzi} A.,  {Charlot} S.,  {Brinchmann} J.,  {White} S.~D.~M.,   {Tremonti}
  C.~A.,  2005, \mn@doi [\mnras] {10.1111/j.1365-2966.2005.09321.x}, \href
  {http://adsabs.harvard.edu/abs/2005MNRAS.362...41G} {362, 41}

\bibitem[\protect\citeauthoryear{{Genzel} et~al.,}{{Genzel}
  et~al.}{2010}]{2010MNRAS.407.2091G}
{Genzel} R.,  et~al., 2010, \mn@doi [\mnras]
  {10.1111/j.1365-2966.2010.16969.x}, \href
  {http://adsabs.harvard.edu/abs/2010MNRAS.407.2091G} {407, 2091}

\bibitem[\protect\citeauthoryear{{Grevesse}, {Asplund}, {Sauval}  \&
  {Scott}}{{Grevesse} et~al.}{2010}]{2010Ap&SS.328..179G}
{Grevesse} N.,  {Asplund} M.,  {Sauval} A.~J.,   {Scott} P.,  2010, \mn@doi
  [\apss] {10.1007/s10509-010-0288-z}, \href
  {http://adsabs.harvard.edu/abs/2010Ap%26SS.328..179G} {328, 179}

\bibitem[\protect\citeauthoryear{{Ho}, {Kudritzki}, {Kewley}, {Zahid},
  {Dopita}, {Bresolin}  \& {Rupke}}{{Ho} et~al.}{2015}]{2015MNRAS.448.2030H}
{Ho} I.-T.,  {Kudritzki} R.-P.,  {Kewley} L.~J.,  {Zahid} H.~J.,  {Dopita}
  M.~A.,  {Bresolin} F.,   {Rupke} D.~S.~N.,  2015, \mn@doi [\mnras]
  {10.1093/mnras/stv067}, \href
  {http://adsabs.harvard.edu/abs/2015MNRAS.448.2030H} {448, 2030}

\bibitem[\protect\citeauthoryear{{Jones}, {Ellis}, {Richard}  \&
  {Jullo}}{{Jones} et~al.}{2013}]{2013ApJ...765...48J}
{Jones} T.,  {Ellis} R.~S.,  {Richard} J.,   {Jullo} E.,  2013, \mn@doi [\apj]
  {10.1088/0004-637X/765/1/48}, \href
  {http://adsabs.harvard.edu/abs/2013ApJ...765...48J} {765, 48}

\bibitem[\protect\citeauthoryear{{Kauffmann} et~al.,}{{Kauffmann}
  et~al.}{2003}]{2003MNRAS.341...33K}
{Kauffmann} G.,  et~al., 2003, \mn@doi [\mnras]
  {10.1046/j.1365-8711.2003.06291.x}, \href
  {http://adsabs.harvard.edu/abs/2003MNRAS.341...33K} {341, 33}

\bibitem[\protect\citeauthoryear{{Kennicutt}}{{Kennicutt}}{1998a}]{1998ARA&A..36..189K}
{Kennicutt} Jr. R.~C.,  1998a, \mn@doi [\araa]
  {10.1146/annurev.astro.36.1.189}, \href
  {http://adsabs.harvard.edu/abs/1998ARA%26A..36..189K} {36, 189}

\bibitem[\protect\citeauthoryear{{Kennicutt}}{{Kennicutt}}{1998b}]{1998ApJ...498..541K}
{Kennicutt} Jr. R.~C.,  1998b, \mn@doi [\apj] {10.1086/305588}, \href
  {http://adsabs.harvard.edu/abs/1998ApJ...498..541K} {498, 541}

\bibitem[\protect\citeauthoryear{{Kewley} \& {Dopita}}{{Kewley} \&
  {Dopita}}{2002}]{2002ApJS..142...35K}
{Kewley} L.~J.,  {Dopita} M.~A.,  2002, \mn@doi [\apjs] {10.1086/341326}, \href
  {http://adsabs.harvard.edu/abs/2002ApJS..142...35K} {142, 35}

\bibitem[\protect\citeauthoryear{{Kewley} \& {Ellison}}{{Kewley} \&
  {Ellison}}{2008}]{2008ApJ...681.1183K}
{Kewley} L.~J.,  {Ellison} S.~L.,  2008, \mn@doi [\apj] {10.1086/587500}, \href
  {http://adsabs.harvard.edu/abs/2008ApJ...681.1183K} {681, 1183}

\bibitem[\protect\citeauthoryear{{Kewley}, {Dopita}, {Sutherland}, {Heisler}
  \& {Trevena}}{{Kewley} et~al.}{2001}]{2001ApJ...556..121K}
{Kewley} L.~J.,  {Dopita} M.~A.,  {Sutherland} R.~S.,  {Heisler} C.~A.,
  {Trevena} J.,  2001, \mn@doi [\apj] {10.1086/321545}, \href
  {http://adsabs.harvard.edu/abs/2001ApJ...556..121K} {556, 121}

\bibitem[\protect\citeauthoryear{{Kewley}, {Dopita}, {Leitherer}, {Dav{\'e}},
  {Yuan}, {Allen}, {Groves}  \& {Sutherland}}{{Kewley}
  et~al.}{2013}]{2013ApJ...774..100K}
{Kewley} L.~J.,  {Dopita} M.~A.,  {Leitherer} C.,  {Dav{\'e}} R.,  {Yuan} T.,
  {Allen} M.,  {Groves} B.,   {Sutherland} R.,  2013, \mn@doi [\apj]
  {10.1088/0004-637X/774/2/100}, \href
  {http://adsabs.harvard.edu/abs/2013ApJ...774..100K} {774, 100}

\bibitem[\protect\citeauthoryear{{Kewley}, {Zahid}, {Geller}, {Dopita}, {Hwang}
   \& {Fabricant}}{{Kewley} et~al.}{2015}]{2015ApJ...812L..20K}
{Kewley} L.~J.,  {Zahid} H.~J.,  {Geller} M.~J.,  {Dopita} M.~A.,  {Hwang}
  H.~S.,   {Fabricant} D.,  2015, \mn@doi [\apjl]
  {10.1088/2041-8205/812/2/L20}, \href
  {http://adsabs.harvard.edu/abs/2015ApJ...812L..20K} {812, L20}

\bibitem[\protect\citeauthoryear{{Leethochawalit}, {Jones}, {Ellis}, {Stark},
  {Richard}, {Zitrin}  \& {Auger}}{{Leethochawalit}
  et~al.}{2016}]{2016ApJ...820...84L}
{Leethochawalit} N.,  {Jones} T.~A.,  {Ellis} R.~S.,  {Stark} D.~P.,  {Richard}
  J.,  {Zitrin} A.,   {Auger} M.,  2016, \mn@doi [\apj]
  {10.3847/0004-637X/820/2/84}, \href
  {http://adsabs.harvard.edu/abs/2016ApJ...820...84L} {820, 84}

\bibitem[\protect\citeauthoryear{{Lilly}, {Carollo}, {Pipino}, {Renzini}  \&
  {Peng}}{{Lilly} et~al.}{2013}]{2013ApJ...772..119L}
{Lilly} S.~J.,  {Carollo} C.~M.,  {Pipino} A.,  {Renzini} A.,   {Peng} Y.,
  2013, \mn@doi [\apj] {10.1088/0004-637X/772/2/119}, \href
  {http://adsabs.harvard.edu/abs/2013ApJ...772..119L} {772, 119}

\bibitem[\protect\citeauthoryear{{Ma}, {Hopkins}, {Faucher-Gigu{\`e}re},
  {Zolman}, {Muratov}, {Kere{\v s}}  \& {Quataert}}{{Ma}
  et~al.}{2016}]{2016MNRAS.456.2140M}
{Ma} X.,  {Hopkins} P.~F.,  {Faucher-Gigu{\`e}re} C.-A.,  {Zolman} N.,
  {Muratov} A.~L.,  {Kere{\v s}} D.,   {Quataert} E.,  2016, \mn@doi [\mnras]
  {10.1093/mnras/stv2659}, \href
  {http://adsabs.harvard.edu/abs/2016MNRAS.456.2140M} {456, 2140}

\bibitem[\protect\citeauthoryear{{Maiolino} et~al.,}{{Maiolino}
  et~al.}{2008}]{2008A&A...488..463M}
{Maiolino} R.,  et~al., 2008, \mn@doi [\aap] {10.1051/0004-6361:200809678},
  \href {http://adsabs.harvard.edu/abs/2008A%26A...488..463M} {488, 463}

\bibitem[\protect\citeauthoryear{{Makarov}, {Prugniel}, {Terekhova}, {Courtois}
   \& {Vauglin}}{{Makarov} et~al.}{2014}]{2014A&A...570A..13M}
{Makarov} D.,  {Prugniel} P.,  {Terekhova} N.,  {Courtois} H.,   {Vauglin} I.,
  2014, \mn@doi [\aap] {10.1051/0004-6361/201423496}, \href
  {http://adsabs.harvard.edu/abs/2014A%26A...570A..13M} {570, A13}

\bibitem[\protect\citeauthoryear{{Martinsson}, {Verheijen}, {Westfall},
  {Bershady}, {Andersen}  \& {Swaters}}{{Martinsson}
  et~al.}{2013}]{2013A&A...557A.131M}
{Martinsson} T.~P.~K.,  {Verheijen} M.~A.~W.,  {Westfall} K.~B.,  {Bershady}
  M.~A.,  {Andersen} D.~R.,   {Swaters} R.~A.,  2013, \mn@doi [\aap]
  {10.1051/0004-6361/201321390}, \href
  {http://adsabs.harvard.edu/abs/2013A%26A...557A.131M} {557, A131}

\bibitem[\protect\citeauthoryear{{Mast} et~al.,}{{Mast}
  et~al.}{2014}]{2014A&A...561A.129M}
{Mast} D.,  et~al., 2014, \mn@doi [\aap] {10.1051/0004-6361/201321789}, \href
  {http://adsabs.harvard.edu/abs/2014A%26A...561A.129M} {561, A129}

\bibitem[\protect\citeauthoryear{{McGaugh}}{{McGaugh}}{1991}]{1991ApJ...380..140M}
{McGaugh} S.~S.,  1991, \mn@doi [\apj] {10.1086/170569}, \href
  {http://adsabs.harvard.edu/abs/1991ApJ...380..140M} {380, 140}

\bibitem[\protect\citeauthoryear{{Mott}, {Spitoni}  \& {Matteucci}}{{Mott}
  et~al.}{2013}]{2013MNRAS.435.2918M}
{Mott} A.,  {Spitoni} E.,   {Matteucci} F.,  2013, \mn@doi [\mnras]
  {10.1093/mnras/stt1495}, \href
  {http://adsabs.harvard.edu/abs/2013MNRAS.435.2918M} {435, 2918}

\bibitem[\protect\citeauthoryear{{Moustakas}, {Kennicutt}, {Tremonti}, {Dale},
  {Smith}  \& {Calzetti}}{{Moustakas} et~al.}{2010}]{2010ApJS..190..233M}
{Moustakas} J.,  {Kennicutt} Jr. R.~C.,  {Tremonti} C.~A.,  {Dale} D.~A.,
  {Smith} J.-D.~T.,   {Calzetti} D.,  2010, \mn@doi [\apjs]
  {10.1088/0067-0049/190/2/233}, \href
  {http://adsabs.harvard.edu/abs/2010ApJS..190..233M} {190, 233}

\bibitem[\protect\citeauthoryear{{Nelson} et~al.,}{{Nelson}
  et~al.}{2016}]{2016ApJ...817L...9N}
{Nelson} E.~J.,  et~al., 2016, \mn@doi [\apjl] {10.3847/2041-8205/817/1/L9},
  \href {http://adsabs.harvard.edu/abs/2016ApJ...817L...9N} {817, L9}

\bibitem[\protect\citeauthoryear{{Nicholls}, {Dopita}  \&
  {Sutherland}}{{Nicholls} et~al.}{2012}]{2012ApJ...752..148N}
{Nicholls} D.~C.,  {Dopita} M.~A.,   {Sutherland} R.~S.,  2012, \mn@doi [\apj]
  {10.1088/0004-637X/752/2/148}, \href
  {http://adsabs.harvard.edu/abs/2012ApJ...752..148N} {752, 148}

\bibitem[\protect\citeauthoryear{{Noeske} et~al.,}{{Noeske}
  et~al.}{2007}]{2007ApJ...660L..43N}
{Noeske} K.~G.,  et~al., 2007, \mn@doi [\apjl] {10.1086/517926}, \href
  {http://adsabs.harvard.edu/abs/2007ApJ...660L..43N} {660, L43}

\bibitem[\protect\citeauthoryear{{Peng}, {Ho}, {Impey}  \& {Rix}}{{Peng}
  et~al.}{2002}]{2002AJ....124..266P}
{Peng} C.~Y.,  {Ho} L.~C.,  {Impey} C.~D.,   {Rix} H.-W.,  2002, \mn@doi [\aj]
  {10.1086/340952}, \href {http://adsabs.harvard.edu/abs/2002AJ....124..266P}
  {124, 266}

\bibitem[\protect\citeauthoryear{{Querejeta} et~al.,}{{Querejeta}
  et~al.}{2015}]{2015ApJS..219....5Q}
{Querejeta} M.,  et~al., 2015, \mn@doi [\apjs] {10.1088/0067-0049/219/1/5},
  \href {http://adsabs.harvard.edu/abs/2015ApJS..219....5Q} {219, 5}

\bibitem[\protect\citeauthoryear{{Queyrel} et~al.,}{{Queyrel}
  et~al.}{2012}]{2012A&A...539A..93Q}
{Queyrel} J.,  et~al., 2012, \mn@doi [\aap] {10.1051/0004-6361/201117718},
  \href {http://adsabs.harvard.edu/abs/2012A%26A...539A..93Q} {539, A93}

\bibitem[\protect\citeauthoryear{{Rich}, {Torrey}, {Kewley}, {Dopita}  \&
  {Rupke}}{{Rich} et~al.}{2012}]{2012ApJ...753....5R}
{Rich} J.~A.,  {Torrey} P.,  {Kewley} L.~J.,  {Dopita} M.~A.,   {Rupke}
  D.~S.~N.,  2012, \mn@doi [\apj] {10.1088/0004-637X/753/1/5}, \href
  {http://adsabs.harvard.edu/abs/2012ApJ...753....5R} {753, 5}

\bibitem[\protect\citeauthoryear{{Rupke}, {Kewley}  \& {Barnes}}{{Rupke}
  et~al.}{2010}]{2010ApJ...710L.156R}
{Rupke} D.~S.~N.,  {Kewley} L.~J.,   {Barnes} J.~E.,  2010, \mn@doi [\apjl]
  {10.1088/2041-8205/710/2/L156}, \href
  {http://adsabs.harvard.edu/abs/2010ApJ...710L.156R} {710, L156}

\bibitem[\protect\citeauthoryear{{Salpeter}}{{Salpeter}}{1955}]{1955ApJ...121..161S}
{Salpeter} E.~E.,  1955, \mn@doi [\apj] {10.1086/145971}, \href
  {http://adsabs.harvard.edu/abs/1955ApJ...121..161S} {121, 161}

\bibitem[\protect\citeauthoryear{{S{\'a}nchez}, {Rosales-Ortega}, {Kennicutt},
  {Johnson}, {Diaz}, {Pasquali}  \& {Hao}}{{S{\'a}nchez}
  et~al.}{2011}]{2011MNRAS.410..313S}
{S{\'a}nchez} S.~F.,  {Rosales-Ortega} F.~F.,  {Kennicutt} R.~C.,  {Johnson}
  B.~D.,  {Diaz} A.~I.,  {Pasquali} A.,   {Hao} C.~N.,  2011, \mn@doi [\mnras]
  {10.1111/j.1365-2966.2010.17444.x}, \href
  {http://adsabs.harvard.edu/abs/2011MNRAS.410..313S} {410, 313}

\bibitem[\protect\citeauthoryear{{S{\'a}nchez} et~al.,}{{S{\'a}nchez}
  et~al.}{2012}]{2012A&A...538A...8S}
{S{\'a}nchez} S.~F.,  et~al., 2012, \mn@doi [\aap]
  {10.1051/0004-6361/201117353}, \href
  {http://adsabs.harvard.edu/abs/2012A%26A...538A...8S} {538, A8}

\bibitem[\protect\citeauthoryear{{S{\'a}nchez} et~al.,}{{S{\'a}nchez}
  et~al.}{2014}]{2014A&A...563A..49S}
{S{\'a}nchez} S.~F.,  et~al., 2014, \mn@doi [\aap]
  {10.1051/0004-6361/201322343}, \href
  {http://adsabs.harvard.edu/abs/2014A%26A...563A..49S} {563, A49}

\bibitem[\protect\citeauthoryear{{S{\'a}nchez} et~al.,}{{S{\'a}nchez}
  et~al.}{2016}]{2016A&A...594A..36S}
{S{\'a}nchez} S.~F.,  et~al., 2016, \mn@doi [\aap]
  {10.1051/0004-6361/201628661}, \href
  {http://adsabs.harvard.edu/abs/2016A%26A...594A..36S} {594, A36}

\bibitem[\protect\citeauthoryear{{Sanders} et~al.,}{{Sanders}
  et~al.}{2016}]{2016ApJ...816...23S}
{Sanders} R.~L.,  et~al., 2016, \mn@doi [\apj] {10.3847/0004-637X/816/1/23},
  \href {http://adsabs.harvard.edu/abs/2016ApJ...816...23S} {816, 23}

\bibitem[\protect\citeauthoryear{{Shirazi}, {Brinchmann}  \&
  {Rahmati}}{{Shirazi} et~al.}{2014}]{2014ApJ...787..120S}
{Shirazi} M.,  {Brinchmann} J.,   {Rahmati} A.,  2014, \mn@doi [\apj]
  {10.1088/0004-637X/787/2/120}, \href
  {http://adsabs.harvard.edu/abs/2014ApJ...787..120S} {787, 120}

\bibitem[\protect\citeauthoryear{{Stott} et~al.,}{{Stott}
  et~al.}{2014}]{2014MNRAS.443.2695S}
{Stott} J.~P.,  et~al., 2014, \mn@doi [\mnras] {10.1093/mnras/stu1343}, \href
  {http://adsabs.harvard.edu/abs/2014MNRAS.443.2695S} {443, 2695}

\bibitem[\protect\citeauthoryear{{Tacconi} et~al.,}{{Tacconi}
  et~al.}{2013}]{2013ApJ...768...74T}
{Tacconi} L.~J.,  et~al., 2013, \mn@doi [\apj] {10.1088/0004-637X/768/1/74},
  \href {http://adsabs.harvard.edu/abs/2013ApJ...768...74T} {768, 74}

\bibitem[\protect\citeauthoryear{{Tremonti} et~al.,}{{Tremonti}
  et~al.}{2004}]{2004ApJ...613..898T}
{Tremonti} C.~A.,  et~al., 2004, \mn@doi [\apj] {10.1086/423264}, \href
  {http://adsabs.harvard.edu/abs/2004ApJ...613..898T} {613, 898}

\bibitem[\protect\citeauthoryear{{Troncoso} et~al.,}{{Troncoso}
  et~al.}{2014}]{2014A&A...563A..58T}
{Troncoso} P.,  et~al., 2014, \mn@doi [\aap] {10.1051/0004-6361/201322099},
  \href {http://adsabs.harvard.edu/abs/2014A%26A...563A..58T} {563, A58}

\bibitem[\protect\citeauthoryear{{VanderPlas}}{{VanderPlas}}{2014}]{2014arXiv1411.5018V}
{VanderPlas} J.,  2014, preprint, \href
  {http://adsabs.harvard.edu/abs/2014arXiv1411.5018V} {} (\mn@eprint {arXiv}
  {1411.5018})

\bibitem[\protect\citeauthoryear{{Veilleux}, {Cecil}  \&
  {Bland-Hawthorn}}{{Veilleux} et~al.}{2005}]{2005ARA&A..43..769V}
{Veilleux} S.,  {Cecil} G.,   {Bland-Hawthorn} J.,  2005, \mn@doi [\araa]
  {10.1146/annurev.astro.43.072103.150610}, \href
  {http://adsabs.harvard.edu/abs/2005ARA%26A..43..769V} {43, 769}

\bibitem[\protect\citeauthoryear{{Vila-Costas} \& {Edmunds}}{{Vila-Costas} \&
  {Edmunds}}{1992}]{1992MNRAS.259..121V}
{Vila-Costas} M.~B.,  {Edmunds} M.~G.,  1992, \mnras, \href
  {http://adsabs.harvard.edu/abs/1992MNRAS.259..121V} {259, 121}

\bibitem[\protect\citeauthoryear{{Walcher} et~al.,}{{Walcher}
  et~al.}{2014}]{2014A&A...569A...1W}
{Walcher} C.~J.,  et~al., 2014, \mn@doi [\aap] {10.1051/0004-6361/201424198},
  \href {http://adsabs.harvard.edu/abs/2014A%26A...569A...1W} {569, A1}

\bibitem[\protect\citeauthoryear{{Whitaker} et~al.,}{{Whitaker}
  et~al.}{2014}]{2014ApJ...795..104W}
{Whitaker} K.~E.,  et~al., 2014, \mn@doi [\apj] {10.1088/0004-637X/795/2/104},
  \href {http://adsabs.harvard.edu/abs/2014ApJ...795..104W} {795, 104}

\bibitem[\protect\citeauthoryear{{Williams} et~al.,}{{Williams}
  et~al.}{1996}]{1996AJ....112.1335W}
{Williams} R.~E.,  et~al., 1996, \mn@doi [\aj] {10.1086/118105}, \href
  {http://adsabs.harvard.edu/abs/1996AJ....112.1335W} {112, 1335}

\bibitem[\protect\citeauthoryear{{Wuyts} et~al.,}{{Wuyts}
  et~al.}{2016}]{2016ApJ...831..149W}
{Wuyts} S.,  et~al., 2016, \mn@doi [\apj] {10.3847/0004-637X/831/2/149}, \href
  {http://adsabs.harvard.edu/abs/2016ApJ...831..149W} {831, 149}

\bibitem[\protect\citeauthoryear{{York} et~al.,}{{York}
  et~al.}{2000}]{2000AJ....120.1579Y}
{York} D.~G.,  et~al., 2000, \mn@doi [\aj] {10.1086/301513}, \href
  {http://adsabs.harvard.edu/abs/2000AJ....120.1579Y} {120, 1579}

\bibitem[\protect\citeauthoryear{{Yuan}, {Kewley}  \& {Rich}}{{Yuan}
  et~al.}{2013}]{2013ApJ...767..106Y}
{Yuan} T.-T.,  {Kewley} L.~J.,   {Rich} J.,  2013, \mn@doi [\apj]
  {10.1088/0004-637X/767/2/106}, \href
  {http://adsabs.harvard.edu/abs/2013ApJ...767..106Y} {767, 106}

\bibitem[\protect\citeauthoryear{{Zaritsky}, {Kennicutt}  \&
  {Huchra}}{{Zaritsky} et~al.}{1994}]{1994ApJ...420...87Z}
{Zaritsky} D.,  {Kennicutt} Jr. R.~C.,   {Huchra} J.~P.,  1994, \mn@doi [\apj]
  {10.1086/173544}, \href {http://adsabs.harvard.edu/abs/1994ApJ...420...87Z}
  {420, 87}

\bibitem[\protect\citeauthoryear{{de Vaucouleurs}, {de Vaucouleurs}, {Corwin},
  {Buta}, {Paturel}  \& {Fouqu{\'e}}}{{de Vaucouleurs}
  et~al.}{1991}]{1991rc3..book.....D}
{de Vaucouleurs} G.,  {de Vaucouleurs} A.,  {Corwin} Jr. H.~G.,  {Buta} R.~J.,
  {Paturel} G.,   {Fouqu{\'e}} P.,  1991, {Third Reference Catalogue of Bright
  Galaxies. Volume I: Explanations and references. Volume II: Data for galaxies
  between 0$^{h}$ and 12$^{h}$. Volume III: Data for galaxies between 12$^{h}$
  and 24$^{h}$.}

\bibitem[\protect\citeauthoryear{{van der Wel} et~al.,}{{van der Wel}
  et~al.}{2014}]{2014ApJ...788...28V}
{van der Wel} A.,  et~al., 2014, \mn@doi [\apj] {10.1088/0004-637X/788/1/28},
  \href {http://adsabs.harvard.edu/abs/2014ApJ...788...28V} {788, 28}

\makeatother
\end{thebibliography}

%%%%%%%%%%%%%%%%%%%%%%%%%%%%%%%%%%%%%%%%%%%%%%%%%%

%%%%%%%%%%%%%%%%% APPENDICES %%%%%%%%%%%%%%%%%%%%%

\appendix

\section{Model line ratios}\label{sec:model_line_ratios}

\begin{figure*}
\includegraphics[width=\linewidth]{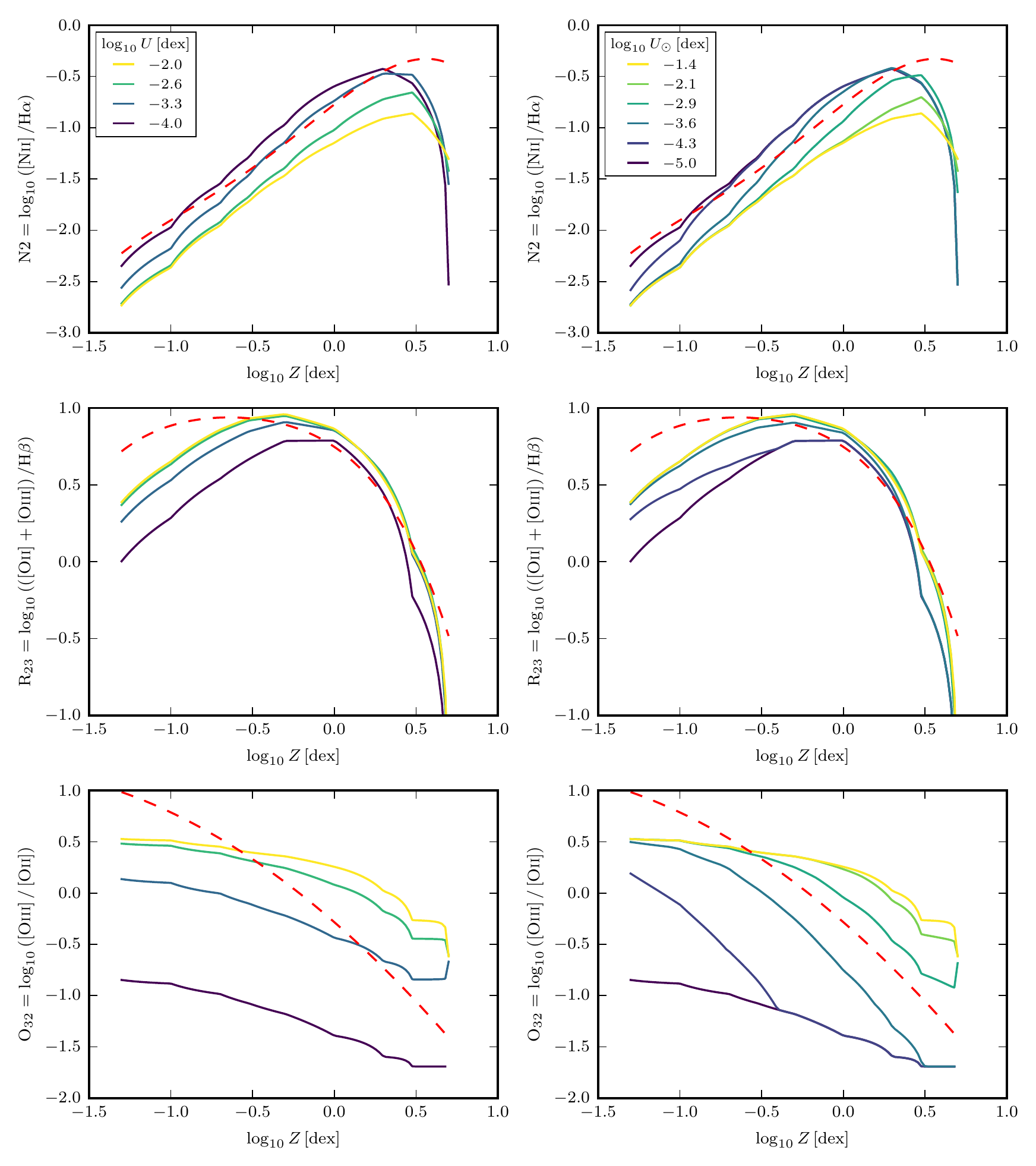}
\caption{Theoretical model predictions for line ratios: $\textrm{N2} = \log_{10}\left( \textrm{\forbidden{N}{ii}{6584}} / \textrm{\Halpha{}} \right)$, $\textrm{R}_{23} = \log_{10}\left( \left( \textrm{\forbidden{O}{ii}{3726,3729}} + \textrm{\forbidden{O}{iii}{4959,5007}} \right) / \textrm{\Hbeta{}} \right)$ and $\textrm{O}_{32} = \log_{10}\left( \textrm{\forbidden{O}{iii}{5007}} / \textrm{\forbidden{O}{ii}{3726,3729}} \right)$.
(Left) We show the \citetalias{2013ApJS..208...10D} models grids with tracks of constant ionization parameter, $\log_{10}U$.
(Right) We show the same model grids, but instead with tracks of constant ionization parameter at solar metallicity, $\log_{10}U_{\sun{}}$, assuming the coupling between metallicity and the ionization parameter (equation~\ref{eq:ionization_parameter_coupling}).
All plots show the parametrizations of \citetalias{2008A&A...488..463M} as a red dashed line.
}
\label{fig:model_line_ratios}
\end{figure*}

In Fig.~\ref{fig:model_line_ratios} we show the \citetalias{2013ApJS..208...10D} model predictions for a set of standard line-ratios.
We show two versions: one with tracks of constant ionization parameter, $\log_{10}U$, and the other with tracks of constant ionization parameter at solar metallicity,  $\log_{10}U_{\sun{}}$.
Both versions span the full model grid range.
To enable this the $\log_{10}U_{\sun{}}$ parameter must span a large range in values $\sim(-5.0,-1.4)$.
As a result the grids are artificially clipped at extreme values of $\log_{10}U_{\sun{}}$.
I.e. at high metallicities the low $\log_{10}U_{\sun{}}$ model tracks pile-up and, vice versa, at low metallicities the high $\log_{10}U_{\sun{}}$ tracks pile-up.
This is most readily seen in $\textrm{O}_{32}$ line-ratio.
In fact there is only a very narrow safe range  $\sim(-3.4,-3.0)$ of $\log_{10}U_{\sun{}}$ values for which there is no clipping at any metallicity.
At first glance this may appear bad, however, no realistic galaxy would span both extremes in metallicity.
The safe range will vary on a galaxy to galaxy basis.
If clipping becomes a significant issue the inferred $\log_{10}U_{\sun{}}$ parameter should become degenerate and unbounded.
Visual inspection of plots, such as those shown in Fig.~\ref{fig:model_fits_0003}, would reveal if clipping has become an issue.

In Fig.~\ref{fig:model_line_ratios} we also compare the model grid predictions with the parametrizations from \citetalias{2008A&A...488..463M}.
We note that there are some discrepancies, especially at low metallicities where the \citetalias{2013ApJS..208...10D} models are unable to reproduce the highest $\textrm{O}_{32}$ values.

It is interesting to also note that in $\textrm{O}_{32}$, which is mostly sensitive to ionization conditions, the \citetalias{2008A&A...488..463M} parametrization shows a similar dependence to the tracks of constant $\log_{10}U_{\sun{}}$.
Much like our approach, \citetalias{2008A&A...488..463M} implicitly encodes some empirical dependence of ionization conditions as a function of metallicity.

\section{Model systematics}\label{sec:model_systematics}

\begin{figure*}
\includegraphics[width=\linewidth]{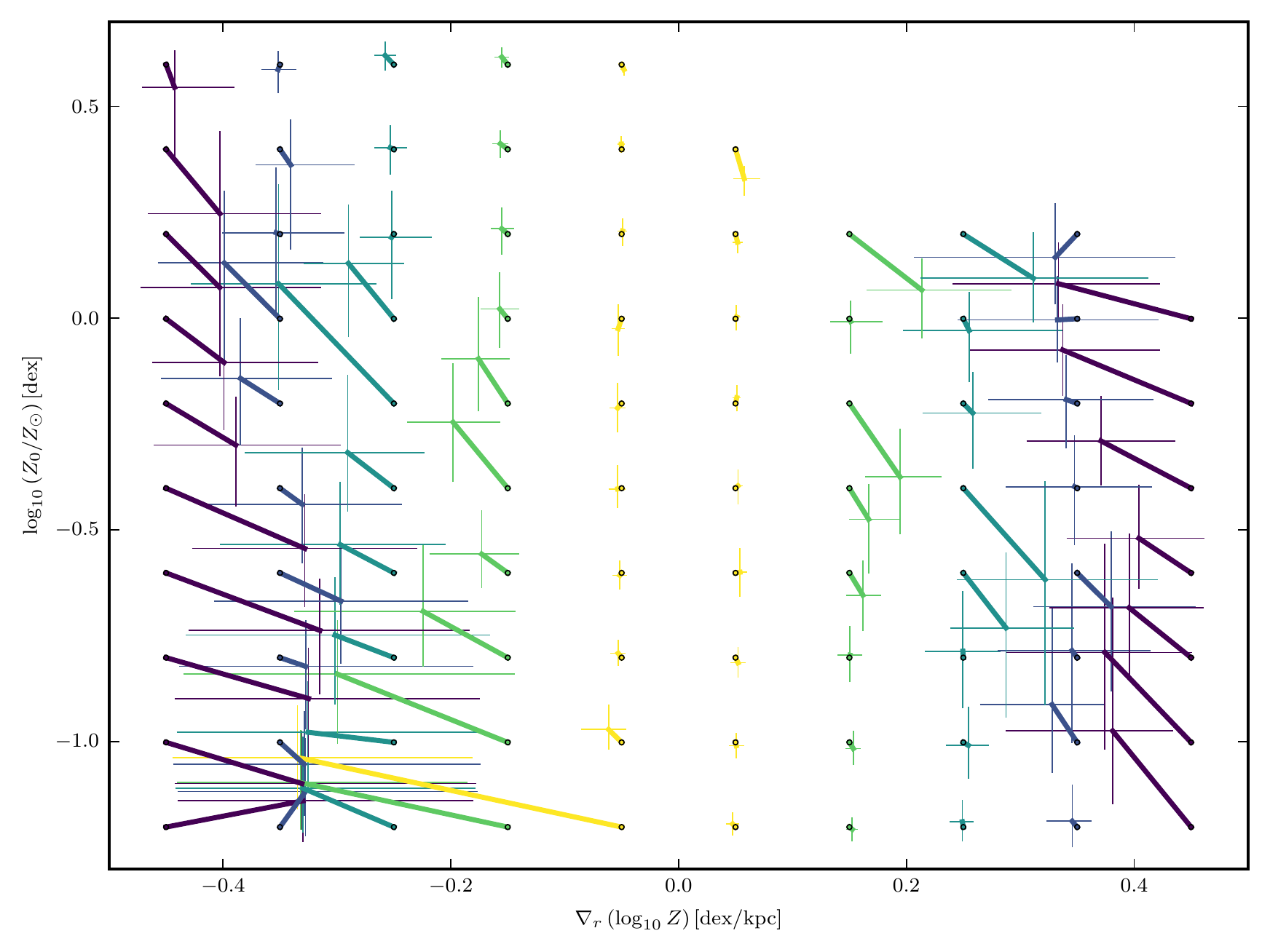}
\caption{Model systematics spanning a wide range of $\log_{10}\left(Z_0/Z_{\sun{}}\right)$ and $\nabla_r\left(\log_{10}Z\right)$ combinations.
The true input model parameters are indicated by circles.
Crosses are plotted at the values inferred by the model.
Size of the crosses indicate the $\pm1\sigma$ errors derived from the 1D marginalized posterior distributions on each parameter.
Thick lines join the crosses to the true value, thereby indicating the systematic offset.
Colours are added primarily to enhance clarity.
Other models input parameters are the same in each model $\textrm{SFR}_\mathrm{tot} = 1\,\textrm{M}_{\sun}\,\mathrm{yr}^{-1}$, $r_d = 0.4\arcsec$, $\log_{10}U_{\sun} = -3\,\textrm{dex}$, $\tau_V = 0.7$, however, they remain free parameters in the fitting.
We use a constant $\textrm{S/N}=6$ as defined on the peak flux of \Hbeta{} line.
At high metallicity \forbidden{O}{ii}{ 3726,3729} and \forbidden{O}{iii}{ 5007} become faint and have insufficient S/N to fit the model.
Therefore models with high central metallicities and steeply positive metallicity gradients (i.e. the upper-right corner) are missing from this plot.
In fact this is in itself an unrelated (but nonetheless important) selection bias on the galaxies we can study.
}
\label{fig:systematics_tests}
\end{figure*}

In Section~\ref{sec:snr_tests} we briefly discussed systematic offsets in the model.
Here we expand upon this by exploring a larger variety of metallicity profiles (i.e. combinations of $\log_{10}\left(Z_0/Z_{\sun{}}\right)$ and $\nabla_r\left(\log_{10}Z\right)$).
This is shown in Fig.~\ref{fig:systematics_tests}, where we fit the model to data generated by the model itself.
The differences that arise indicate systematic offsets.

It can be clearly seen that portions of parameter show strong systematic offsets, typically towards steeper gradients.
However, there is also a distinguishable safe region that runs diagonally from models with high metallicity and negative gradients to models with low metallicity and positive gradients (i.e. from top-left to bottom-right in Fig.~\ref{fig:systematics_tests}).
On the whole models with shallow inferred gradients ($\lvert\nabla_r\left(\log_{10}Z\right)\rvert < 0.2\,\textrm{dex}/\textrm{kpc}$) are free from strong systematics.
However, one cannot truly generalize this statement since this will depend upon, amongst other things, the size of the galaxy, the PSF of the seeing and the S/N of the observations.

There are two related effects that can explain the large systematics we observe.
Firstly we notice that the models with large systematic offsets tend to pileup around \mbox{$\sim\pm0.35\,\textrm{dex}/\textrm{kpc}$} with large errors.
This is highly indicative of model degeneracy and is to be expected since, in the model we clip metallicities to the lower/upper bounds of the \citetalias{2013ApJS..208...10D} model grid.
As a direct result, models with low central metallicities and negative gradients become almost identical.
The same is also true for models with high central metallicities and positive gradients.

The second reason is that we adopt a flat prior on the metallicity gradient, which, as previously noted in Section~\ref{sec:priors}, is not the minimally informative prior.
It is fundamentally harder to distinguish $\nabla_r\left(\log_{10}Z\right) = 0.4$ \& $0.5\,\textrm{dex}/\textrm{kpc}$ models than it to distinguish $\nabla_r\left(\log_{10}Z\right) = 0.1$ \& $0.2\,\textrm{dex}/\textrm{kpc}$ models.
This is true even in the absence of the aforementioned clipping issue, and this should be reflected in the prior by down-weighting steeper gradients.
By choosing a broad, flat prior that includes unrealistic extreme metallicity gradients, we exacerbate the systematics.

A way to partially resolve the issue of systematic errors could be to adopt a joint prior on $\log_{10}Z_0$ and $\nabla_r\left(\log_{10}Z\right)$ which traces the safe region, effectively eliminating the problematic portions of the parameter space.
This of course makes explicit assumptions about the nature of metallicity gradients, but it would formalize such assumptions in a tractable manner.

To summarize the origin of the systematic errors stem from the finite extent of the \citetalias{2013ApJS..208...10D} model grids.
When the model infers galaxies to have extreme metallicity gradients, these should be treated with scepticism.
Investigation of plots such as Fig.~\ref{fig:model_fits_0003} will reveal if a the metallicity gradient is poorly constrained and unbounded.
Overall, one must be acutely aware of the tendency of the model to be biased towards steeper gradients.
However, a careful choice in priors may be able to mitigate against the systematics.

\section{Spaxel binning algorithm}\label{sec:binning_algorithm}

Here we outline our binning algorithm for aggregating spaxels such that the coadded spectrum meets certain acceptance criteria.
In this work our S/N will be defined such that the set of emission-line fluxes are all above a minimum S/N threshold.

Any form of binning trades spatial information for an increased S/N.
This algorithm is intended to reduce the impact of radial information loss, while extracting as many bins as possible, out to large radii.
We therefore need to know what is the galactocentric radius of each spaxel.
With all our data we have higher-resolution images that provides us with accurate estimates for the centre of the galaxy, inclination of the galaxy, and its position angle on the sky.
This inclination is, however, not a good match to the lower-resolution data we are binning.
We use \textsc{Galfit} to fit a 2D Gaussian function to a narrow-band image of a Balmer-series emission line.
We fix the galaxy centre and PA to that of the high-resolution imaging, and obtain the axis ratio of the narrow-band image.
Using these four parameters, we assign radial, $r_i$, and azimuthal coordinates, $\theta_i$. to each spaxel.

The binning algorithm is as follows:
\begin{enumerate}
\item Loop over all spaxels individually.
Perform spectral fitting on each.
If the spaxel's S/N is above the set threshold, assign it a unqiue bin ID number remove spaxel from future binning.

\item \label{item:bin_search} For each remaining unbinned spaxel, coadd the spaxel with other spaxels within $\Delta r$ and $\Delta \theta$ of the spaxel's coordinates.
($\Delta r$ and $\Delta \theta$ define some initial bin size in radial coordinates.)
Perform spectral fitting on the coadded spectrum and record the S/N of the weakest emission line in this bin.

\item \label{item:min_snr}
Find the bin with the lowest S/N, but still above the S/N threshold.
Assign these spaxels with a bin ID number, and remove them from future binning.

\item Repeat steps \ref{item:bin_search} \& \ref{item:min_snr} until there are no bins above threshold.

\item Increase $\Delta r$ and/or $\Delta \theta$ (i.e. increase bin size) and goto step \ref{item:bin_search}.
These increases follow some predefined sequence.
Once $\Delta r$ and/or $\Delta \theta$ reach a maximum size limit, continue to next step.

\item For each remaining unbinned spaxel.
Accrete the spaxel to the nearest bin at a greater radius than it.
If the S/N of the new bin is greater than previous then record the new bin.
Otherwise discard the spaxel and leave the bin unchanged.
\end{enumerate}

\section{SFR Maps}\label{sec:stochmod}

In order to fit our model to the emission-line data we require the SFR distribution as an input.
We could simply fit our data with an exponentially declining SFR density (see equation~\ref{eq:exp_disc}), but, as discussed in section~\ref{sec:real_tests}, clumpy star formation can affect the inferred metallicity profile.
For this reason, we wish to input a best-guess SFR map.

To generate these high-resolution SFR maps we use a combination of multi-band\footnote{F300W, F450W, F606W and F814W} HST imaging and stellar population synthesis (SPS) modelling.
Maps of the SFR and other derived quantities will be published by Shirazi et al. (in prep.).
The modelling procedure is described in detail by \citet{2003MNRAS.341...33K} and \citet{2005MNRAS.362...41G}.
For the SPS models we adopt a star formation history which is a combination of an exponentially declining SFR and superimposed random bursts.
The photometry is calculated using the \citet{2003MNRAS.344.1000B} stellar template library.
Stellar fluxes are attenuated by dust, with the adopted attenuation curve depending on the stellar age.
Young stars ($<10\,\textrm{Myr}$) are attenuated by a $\tau(\lambda) \propto \lambda^{-1.3}$ power-law, whilst older stars will be attenuated by a shallower $\tau(\lambda) \propto \lambda^{-0.7}$ power-law.
This dust model was proposed by \citet{2000ApJ...539..718C}.

For a reliable SPS analysis we require a minimum $S/N\geq5$ in the ($\textrm{F450W} - \textrm{F606W}$) colour image.
To reach this we bin the data using the weighted Voronoi tessellation by \citet{2006MNRAS.368..497D}, a generalization of the algorithm by \citet{2003MNRAS.342..345C}.
Using the SPS modelling we calculate the total SFR in each bin.
However, we wish to partially restore the resolution lost by binning.
We therefore redistribute the binned SFR into the individual pixels using the same proportions as the pixel F814W flux.

Following this procedure we can use high-resolution photometry to produce SFR maps.
We will use these maps as inputs for our emission-line modelling.

\section{Additional plots}

\begin{figure}
\includegraphics[width=\linewidth]{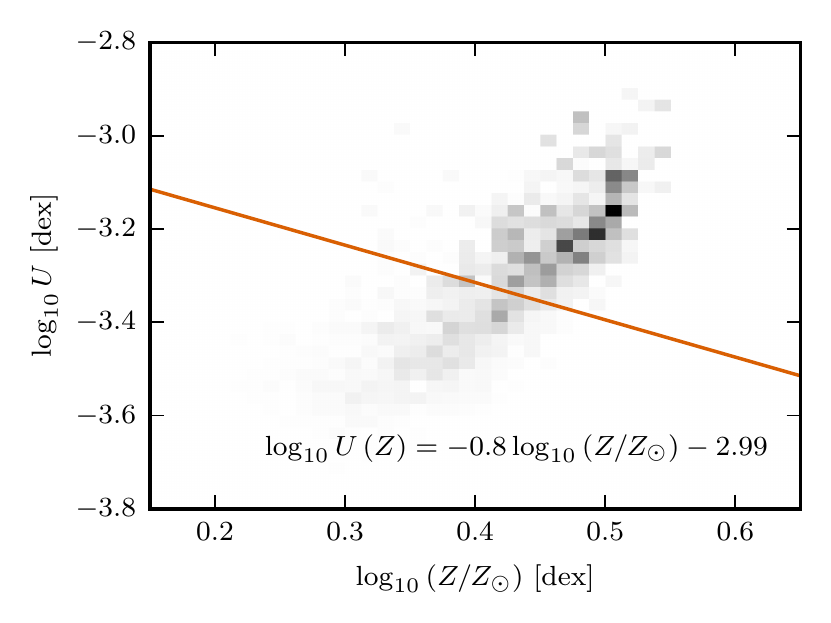}
\caption{Correlation between ionization parameter and metallicity for UGC463.
The individual spaxels are shown as a grey histogram, weighted by the \Halpha{} flux of each spaxel.
The orange line indicates the best fit solution for the $\log_{10}U_{\sun{}}$ assuming the fixed coupling between the ionization parameter and metallicity (i.e. equation~\ref{eq:ionization_parameter_coupling}).
}
\label{fig:UGC463_logZlogU}
\end{figure}

In Fig.~\ref{fig:UGC463_logZlogU} we show the correlation between ionization parameter and metallicity in UGC463.
This positive correlation shows a very different dependence from the typical anti-correlation that we assume.

\begin{figure*}
\includegraphics[width=\linewidth]{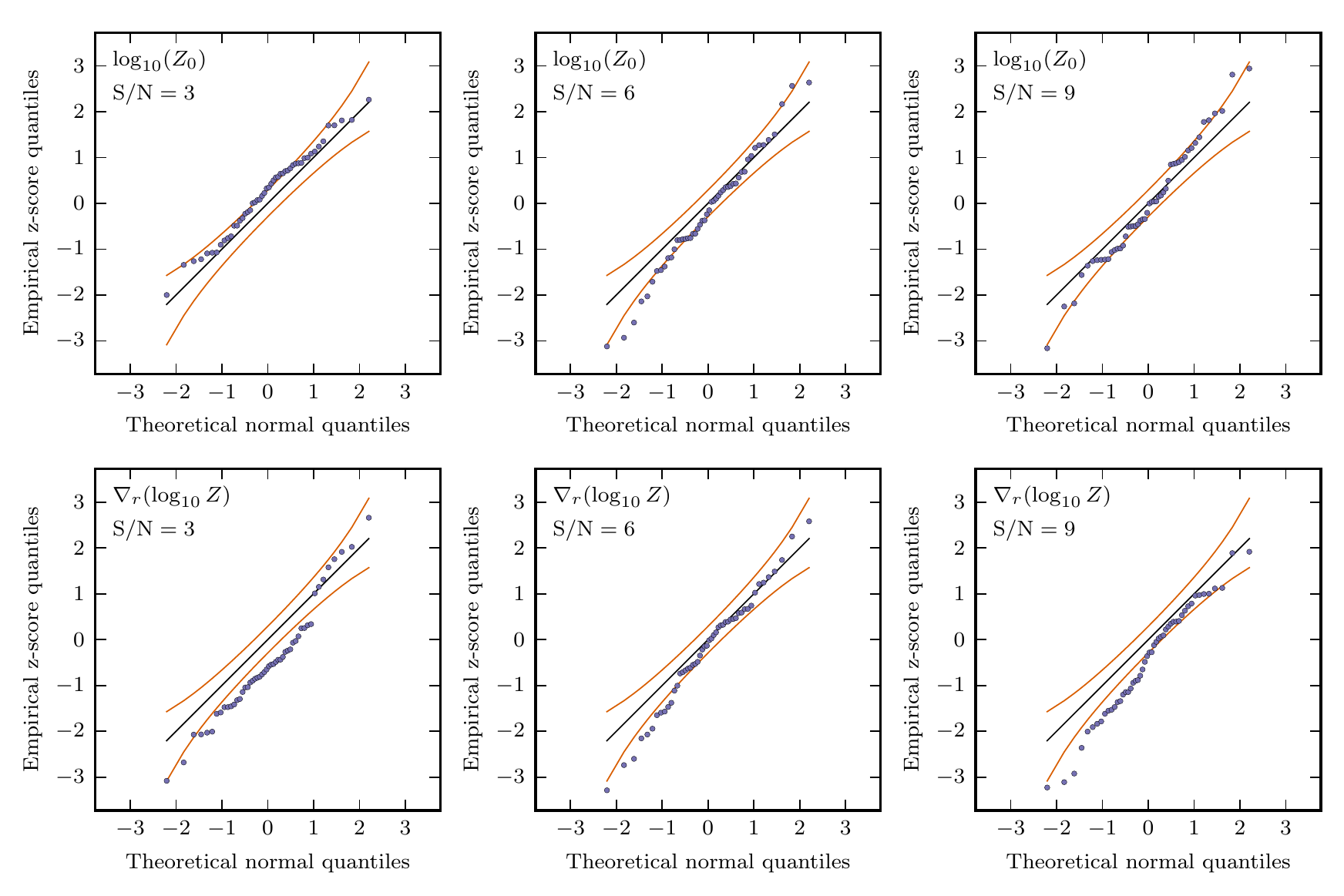}
\caption{Quantile-Quantile plots showing the distribution of inferred model solutions relative to the true input value.
The z-scores of the 50 realizations are plotted on the y-axis, whilst the x-axis shows the z-scores if they were normally distributed.
The orange lines indicate a 90\% confidence interval.
}
\label{fig:QQ}
\end{figure*}

In Fig.~\ref{fig:QQ} we show Quantile-Quantile (Q-Q) for models shown in Tables~\ref{tab:error_test_logZ0}~\&~\ref{tab:error_test_dlogZ}.
If there are no systematic offsets then the data should pass through the (0,0) coordinate (within error).
If the model errors are normally distributed then they should match the black one-to-one line.

%%%%%%%%%%%%%%%%%%%%%%%%%%%%%%%%%%%%%%%%%%%%%%%%%%

% Don't change these lines
\bsp	% typesetting comment
\label{lastpage}
\end{document}